\documentclass[iop]{emulateapj}
\usepackage{color}
\usepackage{graphicx,amsmath,amsfonts,amssymb}
\usepackage[plainpages=false, colorlinks=true, anchorcolor=blue, linkcolor=blue, citecolor=blue, bookmarks=false]{hyperref}

\newcommand{\Msun}{{\rm\,M_\odot}}
\newcommand{\kpc}{{\rm\,kpc}}
\newcommand{\Mpc}{{\rm\,Mpc}}
\newcommand{\Gyr}{{\rm\,Gyr}}
\newcommand{\kms}{{\rm\,km\,s^{-1}}}

\newcommand{\vz}{|\langle{v_{z}}\rangle|}

\newcommand{\fourier}{F_{2}/F_{0}}
\newcommand{\lz}{{L_{z}}}
\newcommand{\normstar}{\widetilde{M}_{\star}}
\newcommand{\normdm}{\widetilde{M}_\text{\rm{DM}}}
\newcommand{\TFC}{\rm{TF_{C}}}
\newcommand{\TFS}{\rm{TF_{G}}}

\defcitealias{kwak17}{Paper I}

\begin{document}

\title{Origin of Non-axisymmetric Features of Virgo Cluster Early-type Dwarf Galaxies. II. Tidal Effects on Disk Features and Stability}

\shorttitle{Tidal Effects on Dwarf Galaxies}
\shortauthors{Kwak et al.}

\author{SungWon Kwak\altaffilmark{1,2}, Woong-Tae
  Kim\altaffilmark{2,3}, Soo-Chang Rey\altaffilmark{4}\& Thomas R. Quinn\altaffilmark{5}}

\affil{$^1$Korea Astronomy and Space Science Institute, Daejeon 305-348, Korea}
\affil{$^2$Department of Physics \& Astronomy, Seoul National University, Seoul 08826, Korea}
\affil{$^3$Center for Theoretical Physics (CTP), Seoul National University, Seoul 08826, Korea}
\affil{$^4$Department of Astronomy and Space Science, Chungnam National University, Daejeon 305-764, Korea}
\affil{$^5$Department of Astronomy, University of Washington, Seattle, WA 98195, USA}

\email{kwakcosmo@gmail.com, wkim@astro.snu.ac.kr, screy@cnu.ac.kr, trq@astro.washington.edu}

\slugcomment{Accepted for publication in the ApJ}

\begin{abstract}
A fraction of dwarf galaxies in the Virgo cluster contain disk features like bars and spiral arms. Using $N$-body simulations, we investigate the effects of tidal forces on the formation of such disk features in disk dwarf galaxies resembling VCC856. We consider 8 Cluster-Galaxy models in which disk dwarf galaxies with differing pericenter distance and spin orientation experience the tidal gravitational force of a Virgo-like NFW halo, and additional 8 Galaxy-Galaxy models in which two dwarf galaxies undergo tidal interactions with different strength.
We find that the cluster tidal effect is moderate due to the small galaxy size, making the bars form earlier by $\sim1$--$1.5\Gyr$ compared to the cases in isolation.
While the galactic halos significantly lose their mass within the virial radius due to the cluster tidal force, the mass of the stellar disks is nearly unchanged, suggesting that the inner regions of a disk-halo system is secured from the tidal force. The tidal forcing from either the cluster potential or a companion galaxy triggers the formation of two-armed spirals at early time before a bar develops. The tidally-driven arms decay and wind with time, suggesting that they are kinematic density waves. In terms of the strength and pitch angle, the faint arms in VCC856 are best matched with the arms in a marginally unstable galaxy produced by a distant tidal encounter with its neighbor $\sim0.85\Gyr$ ago.
\end{abstract}


\section{Introduction}\label{s:intro}
Early-type dwarf galaxies are the most numerous population of galaxies in galaxy clusters, and thus studying their properties is essential to understand the evolutionary history of clusters and galaxy populations therein.
Although dwarf elliptical (dE) galaxies are small in size and known to be dynamically inactive, their kinematical and morphological characteristics vary widely, with some even possessing rotation and hidden disk features \citep{lisker06a,toloba11,janz12}. In particular, a statistical study of \citet{lisker06a} on the disk features of 476 dEs in the Virgo cluster showed that $\sim10\%$ of them actually contain a disk substructure, and the fraction of such a subpopulation classified as dEdis increases to $\sim50\%$ with their luminosity. Due to the disk features and kinematics of dEdis in clusters, it has been inferred that the progenitors of dEdis are infalling late-type dwarf galaxies that have undergone tidal interactions and/or ram-pressure stripping in cluster environments \citep{jerjen00, barazza02, barazza03, simien02, derijcke03, geha03, lisker06a, lisker06b, lisker07, toloba11, toloba12, toloba14, janz12, janz14}.
Also, dEdis were found to be a population of genuine disk galaxies instead of dwarf spheroidal galaxies with a disk component \citep{lisker06a}.
Discoveries of significant rotation in dEdis also support the idea that they indeed originate from disk-type galaxies \citep{toloba11, toloba15}.

By inspecting density profiles of the dEdis in the Virgo cluster, \citet{janz12, janz14} found that the barred and lens populations account for about a half of the total, and some dEdis are located even far from the cluster center. Since the cluster tidal field as a trigger for bar formation is only effective near the cluster core \citep{lokas16}, the origin of the barred dEs
at the outskirts of the cluster was uncertain. To address the bar formation in dEs, in \citet[hereafter \citetalias{kwak17}]{kwak17} we ran $N$-body simulations of 15 dEdis consisting of disk and halo, resembling an observed infalling progenitor of disk dwarf galaxies.
After 10 Gyrs of evolution in isolation, 13 out of the 15 models form a bar, while two models with excessively concentrated halo or a hot disk remain stable without forming a bar. This implies that dEdis are intrinsically unstable to bar formation even without external forces, responsible for the presence of the barred dwarfs at the outskirts of the Virgo cluster found by \citet{janz12,janz14} where the tidal effect is insignificant.
\citetalias{kwak17} also found that the bar-forming models undergo vertical buckling instabilities that thicken the disk vertically, while shortening the bars \citep{combes81,combes90,raha91,merritt94, martinez04}.

Dynamical instabilities of self-gravitating, rotating stellar disks are usually invoked as a formation channel of galactic bars \citep{miller70,hohl71,kalnajs72,kalnajs77}.
The bar formation time and physical properties in normal disk galaxies depend on various galaxy parameters including the mass and concentration of dark matter halo \citep{ostriker73,christodoulou95,sellwood01}, disk scale height \citep{klypin09}, fraction of counter-streaming stars \citep{sellwood94}, degree of radial random motions \citep{athanassoula86}, and presence of the gaseous component \citep{berentzen07,athanassoula13,seo19}. Bars form more easily in galaxies with a smaller dark halo concentration, a thinner and colder disk, a lower fraction of the counter-streaming stars, and/or less gas.
\citetalias{kwak17} showed that the effects of these parameters on bars in dEdis are qualitatively similar to those in normal disk galaxies.

In addition to the internal properties of galaxies, tidal forces can
independently trigger the bar formation. Most previous studies considered galaxy-galaxy encounters and found that the bar strength and pattern speed are correlated with the impact parameter of the galaxy encounters. (e.g., \citealt{gerin90,noguchi96,miwa98,oh08,berentzen04,lang14,lokas14,oh15,gajda17,lokas18}). Tidal forcing can also be provided by the cluster potential itself for galaxies in orbital motions, potentially reshaping their morphology and properties.  Owing to the strong tidal field, the outer region of extended galactic halos beyond their tidal radii can be vigorously truncated \citep{richstone76,white76,merritt83}. \citet{moore96,moore98} showed that the combined effects of multiple fast encounters as well as the cluster tidal field are able to transform late-type galaxies into dEs or dwarf spheroidals. Hierarchical cluster formation, in which the cluster mass grows in time, may weaken the tidal effects \citep{moore99,gnedin03a,gnedin03b}, but the cluster tidal field and galaxy encounters still play an important role in the galaxy metamorphosis, possibly destroying low luminosity galaxies with relatively large scale lengths.
\citet{smith10} investigated the effects of harassment on infalling late-type galaxies and showed that strong transformations may occur in cluster cores, albeit infrequently. Yet, the effects of the cluster tidal force on the properties of non-axisymmetric features in dEdis has to be explored.

The existence of two-armed spirals has occasionally been reported among the Virgo dEdis, notably VCC856, suggesting that such dEdis with spirals are under rotation \citep{jerjen00,lisker06a,lisker09,lisker09b}. These arms are so faint that they can be recognized only after unsharp masking. One of the widely accepted mechanisms for the formation of spiral arms is swing amplifications of inherent noises or internal perturbations by molecular clouds in galactic disks, usually resulting in multiple arms that are transient and recurrent (e.g., \citealt{fujii11,grand12,grand13,baba13}). When disk galaxies undergo tidal interactions which induce $m=2$ perturbations, however, swing amplifications produce two arms close to kinematic density waves, modified by self-gravity, in disks with a central bulge component that tends to suppress the bar instability triggered by the tidal encounters (e.g., \citealt{donner94, oh08, oh15,pettitt16,semczuk17}).
Although the two-armed spiral structures in VCC856 are driven most likely by tidal forcing, it is questionable which tidal force between cluster tidal field and a galaxy-galaxy encounter is more important for the arm formation.
Since VCC856 is a bulgeless galaxy that is usually unstable to bar formation, a weak tidal force should be applied before the bar formation to account for the faint spirals in VCC856. Previous work that studied arm formation by galaxy-galaxy encounter (e.g., \citealt{donner94, oh08, oh15,pettitt16}) or by the cluster tidal field (e.g., \citealt{byrd90,semczuk17}) all considered normal disk galaxies, so that it is interesting to study how effective the tidal forces will be in inducing spiral structure in dEdis.

In this paper, we run $N$-body simulations of dEdis to investigate how the cluster tidal field and galaxy interactions separately alter their dynamical evolution in galaxy clusters. This extends \citetalias{kwak17} in which they were evolved in isolation. We borrow two galaxy models from \citetalias{kwak17}, representing either stable or unstable model of VCC856 within observational error ranges, and let them orbit under the cluster potential or undergo a tidal encounter with a neighbor.  We explore how the cluster tidal force affects the stability of stellar disks, the bar shape and strength, vertical buckling instabilities, and the halo mass truncation, in comparison with those in the isolated counterparts. We also quantify the physical properties of spiral arms induced by the tidal forces. By comparing arms in simulations with those in VCC856, we suggest a probable formation mechanism for spiral arms in dEdis.

This paper is organized as follows. In Section \ref{s:ch2}, we introduce our Cluster-Galaxy and Galaxy-Galaxy models and numerical methods. In Section \ref{s:ch3}, we present the properties of early spiral arms and late-time bars produced in the Cluster-Galaxy models. We also describe the dependence of the changes in the galaxy mass and angular momentum transfer on the tidal force and the spin orientation. In Section \ref{s:ch4}, we present the properties of the spiral arms created in the Galaxy-Galaxy models. In Section \ref{s:ch5}, we summarize our results and discuss them in the context of galaxy evolution in cluster environments.

\section{Models and Methods}\label{s:ch2}
To investigate dynamical evolution of dwarf disk galaxies subject to various tidal perturbations, we choose galaxy models constructed in \citetalias{kwak17} and run two types of simulations: Cluster-Galaxy models and Galaxy-Galaxy models. In the Cluster-Galaxy models, a galaxy follows an eccentric orbit about the center of a Virgo-like cluster and responds to the gravitational force of the cluster. In the Galaxy-Galaxy models, two galaxies interact gravitationally on their mutual parabolic orbits.

\subsection{Galaxy Models}
In \citetalias{kwak17}, we chose VCC856, one of dEdis in the Virgo cluster, as our standard model to represent infalling progenitors and constructed 15 isolated galaxy models allowing for uncertainties in its observed properties. Photometric studies show that it is nearly gas free, has an exponential stellar disk without a bulge component, and possesses faint $m=2$ spirals with amplitudes $\sim3$--$4$\% of the disk \citep{jerjen00}. In terms of kinematics, it is rotationally supported \citep{simien02} and is not dominated by a dark halo within the effective radius \citep{toloba14}. These properties suggest that VCC 856 might have experienced ram-pressure stripping in the cluster, but undergone neither strong tidal interactions nor major mergers, preserving the early shapes of infalling late-type dEs. In this paper, we borrow two galaxy models, Galaxies S1 and DM2, from \citetalias{kwak17} and study their dynamical evolution subject to time-varying tidal forces.

Our progenitor galaxy models consist of an exponential stellar disk embedded in a dark matter halo that follows the Hernquist profile \citep{hernquist90}.
Galaxy S1 is the standard model whose properties are the same as the mean observed values of VCC856 in terms of the rotation curve \citep{simien02}, velocity anisotropy \citep{gerssen00,lisker09}, stellar mass and central dark matter fraction \citep{toloba14}, and median axial ratio \citep{lisker07,lisker09}.
It has the total disk mass $M_d=10^9\Msun$, the disk scale length  $r_d=0.625\kpc$, the disk scale height  $z_d=0.33R_d$, the halo mass $M_\text{DM}=1.5\times10^{10}\Msun$, the halo concentration $c=9$, the dark matter fraction  $f_\text{DM}(r<r_{\rm{eff}})\sim30\%$ within the effective radius $r_{\rm{eff}}=1.68 r_{d}$, with $r_d$ being the disk scale radius, and the velocity anisotropy $f_{r}=\sigma_{r}^2/\sigma_{z}^2=1.56$. \citetalias{kwak17} showed that, when evolved in isolation, Galaxy S1 with the Toomre stability parameter of $Q_{T}\sim1.3$ at $r=0.7\kpc$ forms a bar that starts to grow from $t\sim2\Gyr$, achieves a peak strength at $t\sim 6.4\Gyr$, and undergoes an episode of buckling instability at $t\sim6.7\Gyr$. On the other hand, Galaxy DM2 has $c=20$ and $f_\text{DM} (r<r_{\rm{eff}})\sim54\%$, corresponding to the upper limit of the observed dark matter fraction, while the other properties are kept the same as in Galaxy S1. The strong dark matter concentration leads to $Q_T\sim1.8$ at $r=1.7\kpc$, making the disk in Galaxy DM2 stable and featureless until the end of the run ($10\Gyr$).

\begin{deluxetable}{ccllccc}
\tablecaption{Cluster-Galaxy Model Parameters\label{tbl:cluster}}
\tablewidth{0pt}
\tablehead{\colhead{Model}
          & \colhead{Galaxy}
          & \colhead{$(X,Y,Z)_i$}
          & \colhead{$(v_X,v_Y,v_Z)_i$}
          & \colhead{Spin}
          & \colhead{$\rm{TF_{C}}$} \\
            \colhead{$~$}
          & \colhead{}
          & \colhead{[Mpc]}
          & \colhead{[km/s]}
          & \colhead{$~$}
          & \colhead{$~$} \\
            \colhead{(1)}
          & \colhead{(2)}
          & \colhead{(3)}
          & \colhead{(4)}
          & \colhead{(5)}
          & \colhead{(6)}}
\startdata
C1 & S1  & $(-0.5,0,0)$  & $(0,-350,0)$ & pro- & $0.26$ \\
C2 & S1  & $(-0.75,0,0)$ & $(0,-350,0)$ & pro- & $0.16$ \\
C3 & S1  & $(-1.0,0,0)$  & $(0,-350,0)$ & pro- & $0.11$ \\
C4 & S1  & $(-1.5,0,0)$  & $(0,-350,0)$ & pro- & $0.060$ \\
C5 & DM2 & $(-0.5,0,0)$  & $(0,-350,0)$ & pro- & $0.16$ \\
C6 & DM2 & $(-1.5,0,0)$  & $(0,-350,0)$ & pro- & $0.037$ \\
R1 & S1  & $(-0.5,0,0)$  & $(0,+350,0)$ & retro- & $0.26$ \\
I1 & S1  & $(-0.5,0,0)$  & $(0,0,-350)$ & ortho- & $0.26$ \\\enddata
\tablecomments{Column (1) is the model name. Column (2) is the galaxy name that orbits around the Virgo-like dark matter halo. Columns (3) and (4) indicate the initial position and velocity of the galaxy. The ratio of apocenter to pericenter is set to 5. Column (5) lists the spin orientation of the galaxy relative to its orbit: `pro-' and `retro-' stand for prograde and retrograde spin, while `ortho-' corresponds to the stellar disk perpendicular to its orbital plane. Column (6) gives the dimensionless tidal force at the pericenter.}
\end{deluxetable}

\subsection{Cluster-Galaxy Models}

Our Cluster-Galaxy models consist of a cluster and a progenitor galaxy that orbits about the cluster center.
For the cluster, we follow the realization procedure of \citet{kazantzidis04} to set up the NFW form of the dark matter distribution
\begin{equation}
   \rho_\text{DM} (R)= \frac{\rho_{0}}{(R/R_s)(1+R/R_s)^2},
\end{equation}
where $\rho_0$ is the characteristic density, $R_s$ is the scale radius, and $R$ is the clustocentric radius \citep{navarro97}. The corresponding virial mass $M_\text{vir}$ and radius $R_\text{vir}$ are given by
\begin{equation}
   M_\text{vir} = 4\pi\rho_{0}R^3_s\bigg[ \text{ln}(1+c_\text{cl})-\frac{c_\text{cl}}{1+c_\text{cl}}\bigg],  \quad R_\text{vir}= c_\text{cl} R_s,
\end{equation}
with the concentration parameter $c_\text{cl}$. Our adopted cluster has $M_{\rm{vir}}=5.4\times10^{14}\Msun$ and the concentration $c_\text{cl}=3.8$, resembling the properties of the Virgo cluster estimated via gravitational lensing \citep{comerford07}. Since the total mass of the NFW distribution is divergent, we employ an exponential cutoff with the scale length of 0.1$R_{\rm{vir}}$ beyond the virial radius $R_{\rm{vir}}=2.1\Mpc$  \citep{springel99}.

We consider eight Cluster-Galaxy models by varying the galaxy type, orbital radius, and spin orientation of galactic disks. Table \ref{tbl:cluster} lists the parameters of each model. In Models C1--C4, we place Galaxy S1 on four orbits with different (clustocentric) apocenter distances $R_\text{apo}=0.5, 0.75, 1.0, 1.5\Mpc$. The initial orbital velocity $(v_X, v_Y, v_Z)_i$ is chosen to make the ratio of apocenter to pericenter distances to $R_\text{apo}/R_\text{peri}=5$ for all models, corresponding to a median value for dark haloes within galaxy clusters in cosmological simulations \citep{ghigna98}. These orbital parameters of Model C1--C4 are taken from \citet{lokas16} who studied tidal responses of normal disk galaxies. In Models C5 and C6, we place Galaxy DM2 on the innermost and outermost orbits, respectively. For Models C1 to C6, the disks spin in the same direction as their orbits that are in the counterclockwise direction seen from the positive $Z$-axis. Models R1 and I1 are the same as Model C1 except for the galaxy orbit.  In Model R1, the orbital angular momentum of the galaxy is inverted relative to the spin orientation. In Model I1, the galaxy orbits about the negative $Y$-axis, so that the tidal force would be exerted in the direction perpendicular to the disk at each pericenter and apocenter.

To quantify the cluster tidal effects on the galaxies, we define the dimensionless tidal force parameter $\TFC$ as
\begin{equation}
{\TFC}= \left(\frac{M_\text{cl}}{M_\text{gal}}\right)
   \left(\frac{r_\text{gal}}{R_\text{peri}}\right)^3,
\end{equation}
where $M_{\rm{cl}}$ is the enclosed mass of the cluster within $R_{\rm{peri}}$, and $M_{\rm{gal}}$ is the total (stars plus dark matter) galaxy mass within $r_{\rm{gal}}=7r_d$ (e.g., \citealt{byrd92}).
Within $r_{\rm{gal}}$, the enclosed disk mass is approximately the same as the total disk mass, but the enclosed halo mass differs in Galaxies S1 and DM2, owing to their different halo concentration: $M_{\rm{gal}}= 2.88\times10^9 \Msun$ for Galaxy S1 and $4.68\times10^9 \Msun$ for Galaxy DM2.
The corresponding $\TFC$ values are listed in Column 6 of Table \ref{tbl:cluster}.

\subsection{Galaxy-Galaxy Models}
We additionally construct the Galaxy-Galaxy models to compare the differences in the properties of the spiral arms induced by a cluster tide and a galaxy encounter.
Each Galaxy-Galaxy model consists of a galaxy and a perturber moving on mutual parabolic, prograde orbits. As a target galaxy, we consider Galaxies S1 and DM2, but use only Galaxy DM2 as a perturbing galaxy for all models. To control tidal forcing, we vary the (galactocentric) pericenter distance $r_\text{peri}$ from $9.6\kpc$ to $16.5\kpc$. The initial separation between the galaxy and perturber is kept fixed to $50\kpc$. These parameters are chosen so as to study the formation of faint spirals in the inner regions of galaxies via tidal encounters. In total, we consider 8 models of galaxy encounters. The model parameters and simulation results are presented in Section \ref{s:ch4}.

\begin{figure*}
\centering\includegraphics[angle=0,width=17cm]{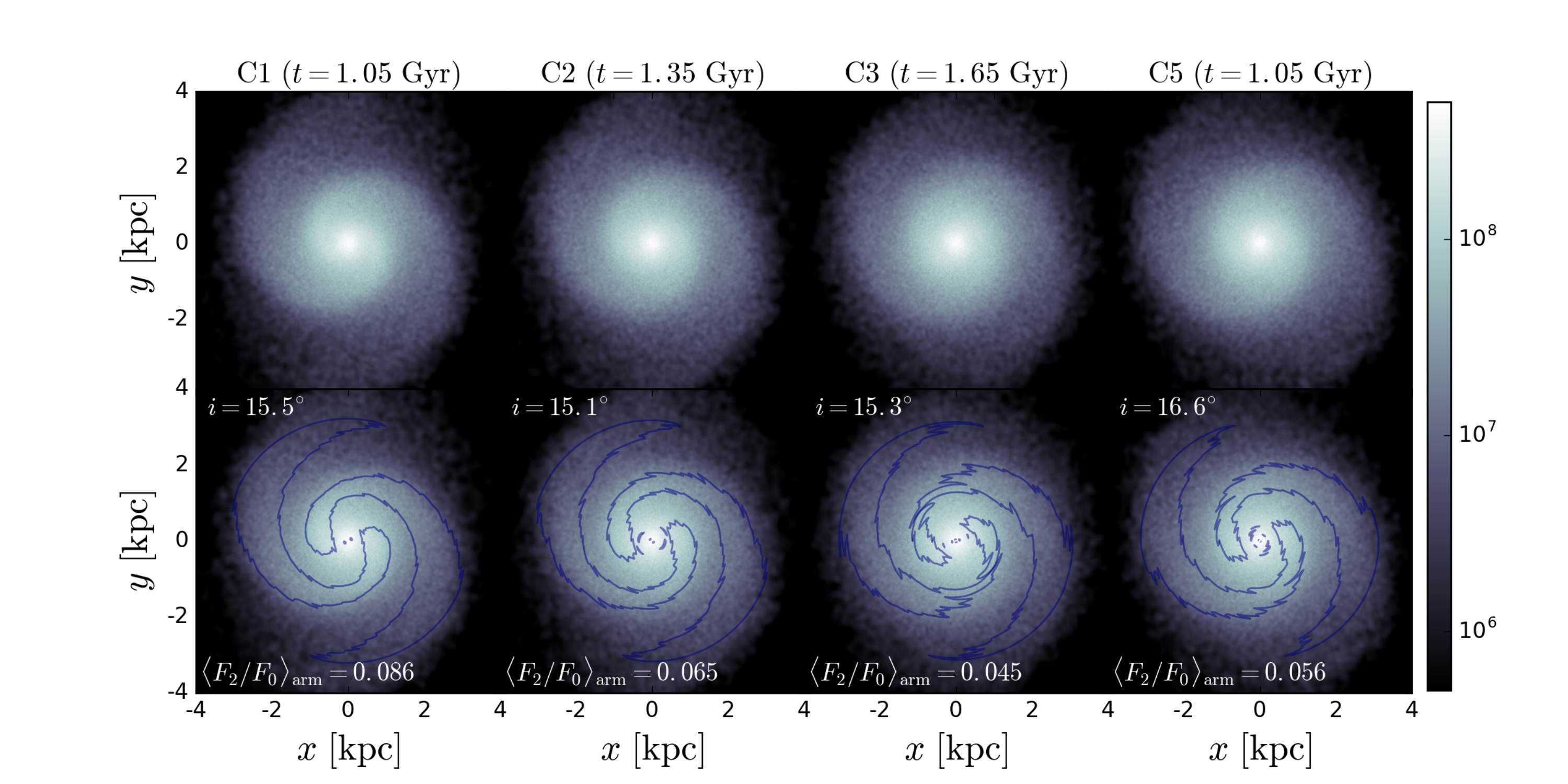}
\caption
{Distributions of the surface density $\Sigma$ of the stellar disk projected along the $z$-direction in Models C1, C2, C3, and C5, from left to right, that possess spiral arms. The times chosen are $0.35\Gyr$ after the first pericenter passage for all models. The colorbar indicates $\Sigma$ in units of $M_\odot {\rm\,kpc}^{-2}$. Overlaid in the bottom panels are the overdensity contours of the $m=2$ Fourier mode. The mean strength $\langle F_{2}/F_{0} \rangle _{\rm{arm}}$ and the pitch angle $i$ of the arms are displayed.}\label{fig01spr}
\end{figure*}

\subsection{Numerical Method}

Using the GALIC code \citep{yurin14}, we construct the initial configurations of the galaxy models as well as the Virgo-like cluster halo (see \citetalias{kwak17}). The numbers of particles distributed are $5\times10^5$ for the stellar disk, $2\times10^6$ for the galactic dark halo, and $10^6$ for the cluster halo. The softening length of each component is set equal to the  mean particle distance within the half mass radius. All of our simulations are run using the \textsc{changa} code, a hybrid $N$-body and SPH code, with the force accuracy option $\theta=0.7$ and the timestepping scale $\eta=0.1$ \citep{jetley08,jetley10,menon15}.
We evolve the Cluster-Galaxy and the Galaxy-Galaxy models for 10 Gyr and 1.5 Gyr, respectively. We analyze and visualize the outputs by using \textsc{pynbody}, a python package specialized for $N$-body and SPH simulations \citep{pynbody}.

\section{Cluster-Galaxy Interaction}\label{s:ch3}

Our galaxies in Models C1 to C6 placed initially at their apocenters $X=-R_{\rm{apo}}$ start to move toward the negative $Y$-direction. They orbit about the cluster center in the counterclockwise direction seen from the positive $Z$-axis, which is in the same sense as the disk spin. In general, the galaxy orbits are not closed in the inertial frame, drawing rosetta shapes. The trajectories of the galaxy orbits, the temporal changes of the clustocentric distance, and orbital velocity of the galaxies in Models C1--C4 are overall similar to those shown in Figures 1 and 2 of \citet{lokas16}.
The galaxy in Model C1 to C4 reaches the pericenter for the first time at $t_{\rm{peri}}=0.7, 1.0, 1.3, 1.9 \Gyr$ with the velocity $v_\text{gal} =1496, 1670, 1782, 1910\kms$, respectively.
The galaxy orbits in Models C5 and C6 are similar to those in Models C1 and C2, respectively. The galaxies in Models R1 and I1 are set to initially move in the positive $Y$- and negative $Z$-directions, respectively.

\citet{donghia2010} demonstrated that in a galaxy-galaxy encounter, a quasi-resonance between galaxy orbit and its spin may result in a strong tidal response in the disk where its spin frequency $\Omega_{\rm{disk}}$ becomes comparable to the orbital angular frequency $\Omega_{\rm{orb}}$ of the perturbing galaxy. The similar resonance may occur in our Cluster-Galaxy models at the radius where $\Omega_{\rm{disk}}=v_\text{cir}/r$ is equal to  $\Omega_{\rm{orb}}= v_\text{gal} /R_\text{gal}$, where $v_{\rm{cir}}$ is the circular velocity of the disk spin and $r$ is the galactocentric distance. The orbits of our galaxies predict that the quasi-resonant radii are $r= 2.80$, 3.41, 3.94, and $4.90 \kpc$ for Models C1 to C4, respectively.  Since dEdis are very small in size, this implies that the tidal response would not be very effective even for Model C1 with strongest tidal forcing, whose effective disk radius is smaller than the quasi-resonant radius by a factor of $\sim2.7$.

\subsection{Early Spiral Arms}\label{sec:sparm}

Weak tidal forces due to the cluster potential form a two-armed spiral in some prograding models at early time. Figure \ref{fig01spr} plots the face-on views of the surface density of the stellar disk in Models C1, C2, C3, and C5 at $t=t_\text{peri}+0.35\Gyr$ before developing a small bar, together with the contours of the $m=2$ Fourier mode. Tidal perturbations in Models C4 and C6 are too weak to produce clear spiral structures in the inner regions of the disks.
Models R1 and I1 with different spin orientations are not susceptible to the formation of spiral arms. After this time, galaxies possess both spirals and bars, the latter of which soon grow to dominate the non-axisymmetric features.

We quantify the arm strength by $\langle F_2/F_0\rangle_\text{arm}$, where $F_m$ denotes the amplitude of the Fourier transform
\begin{equation}\label{eq:fm}
   F_{m}(r) = \sum_{j} \mu_{j}(r) e^{im\phi_j},
\end{equation}
where $\mu_j$ and $\phi_j$ are the mass and azimuthal angle of the $j$-th particle in a radial bin with width $\Delta r=0.1\kpc$ centered at $r$,
and $\langle \;\rangle_\text{arm}$ indicates a spatial average over $r=1.0$--$2.0\kpc$, consistent with the radial range of the observed arms in VCC 856 \citep{jerjen00}. The arm pitch angle $i$ is determined as $i=\rm{arctan} (2/\it{p}_{\rm{max}})$ in the same radial range, where $p_{\rm{max}}$ is the slope in the ln$R$--$\phi$ plane of the locus of the maximum $F_2$. The arm strength and pitch angle measured at $t=t_\text{peri}+0.35\Gyr$ are found to be $\langle\fourier\rangle_{\rm{arm}}=5\sim9\%$ and $i=15.1^\circ$--$16.6^\circ$ for Models C1, C2, C3 and C5, as given in Figure \ref{fig01spr}.

\begin{figure}
\centering\includegraphics[angle=0,width=8.5cm]{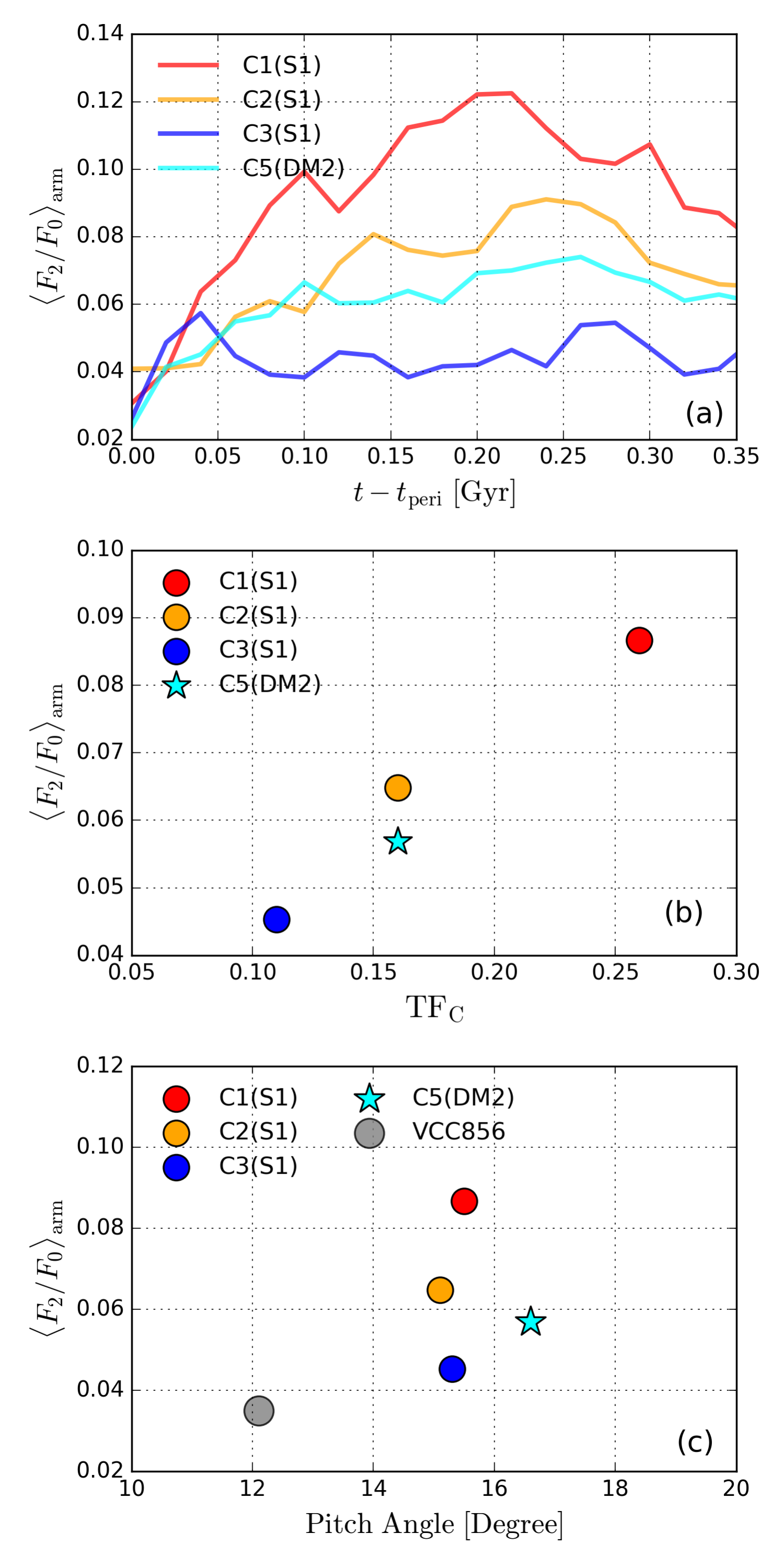}
\caption
{(a) Variations of the arm strength $\langle F_{2}/F_{0} \rangle_\text{arm}$  as a function of time elapsed from the first pericenter passage $t_\text{peri}$ in Models C1, C2, C3, and C5. Dependence of $\langle F_{2}/F_{0}\rangle_\text{arm} $ measured at $t=t_\text{peri}+0.35\Gyr$ on (b) the tidal forcing parameter $\rm{TF_{cls}}$ and (c) the arm  pitch angle. The grey circle in (c) corresponds to the observed arms in VCC856 for comparison.
}\label{fig02spr}
\end{figure}

\begin{figure}
\centering\includegraphics[angle=0,width=8.5cm]{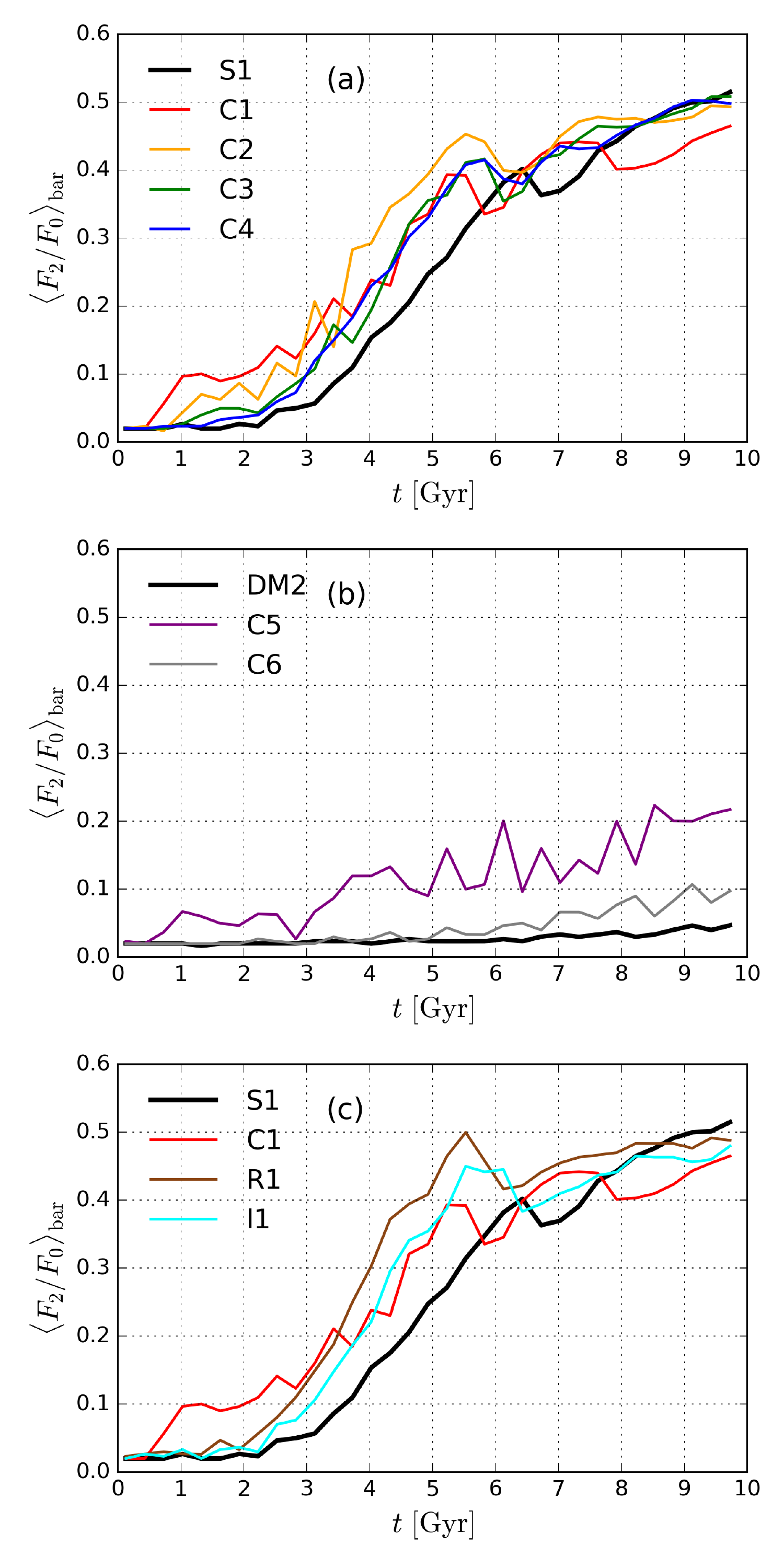}
\caption
{Temporal evolution of the bar strength $\langle F_{2}/F_{0} \rangle_\text{bar}$ averaged over the regions with $r=1.3$--$1.7\kpc$ for (a) Models C1--C4, (b) Models C5 and C6, and (c) Models C1, R1, and I1. In all panels, the results of Galaxy S1 and DM2 in isolation are compared as black solid lines. Note that $\langle F_{2}/F_{0} \rangle_\text{bar}$ at $t
\lesssim 2\Gyr$, in Models C1--C3 and C5, measures the strength of spiral arms that form earlier than the bar.
}\label{fig03fourier}
\end{figure}

Figure \ref{fig02spr} plots the temporal variations of the arm strength after $t=t_\text{peri}$ as well as its correlation with the tidal force parameter $\TFC$ and the arm pitch angle measured at $t=t_\text{peri}+0.35\Gyr$ for models with spiral arms. In the top panel, Model C1, which hosts Galaxy S1 on the innermost orbit, forms the strongest arms, while Model C3 possesses the weakest arms due to largest $R_\text{peri}$. Although Models C2 and C5 have the same $R_\text{peri}$, arms are weaker in Model C5 that hosts Galaxy DM2. This indicates that the central concentration of a dark halo tends to stabilize the stellar disk. Note that the arm strength is almost linearly proportional to the tidal forcing $\TFC$ that allows for central concentration via $M_{\rm{gal}}$. Overall, the arms decay and wind over time after achieving the peak strength, with the decaying rate faster for stronger arms. This suggests that the tidally-driven arms are kinematic density waves (e.g., \citealt{oh08,oh15}).

Figure \ref{fig02spr} shows that the arms formed in our Cluster-Galaxy models are comparable to the faint spirals in VCC856 with $\langle F_2/F_0\rangle_{\rm{arm}}=0.03$--$0.04$ and $i=12.1^\circ$ \citep{jerjen00}, although the former is slightly less wound than the latter. Presumably, a model with tidal forcing slightly weaker than Model C3 (but stronger than Model C4, i.e., $0.06\lesssim \TFC\lesssim 0.11$) may produce spiral arms comparable, in strength, to those in VCC856. The simulated arms, however, do not wind out below 15$^\circ$ at $t=t_\text{peri}+0.35\Gyr$ after which a small bar begins to grow at $r\lesssim 1\kpc$. In Section \ref{s:ch4}, we shall show the formation of the spiral arms induced by a galaxy interaction.

\subsection{Late-time Bar}
\subsubsection{Bar Formation and Buckling Instability}

\begin{figure*}
\centering\includegraphics[angle=0,width=17cm]{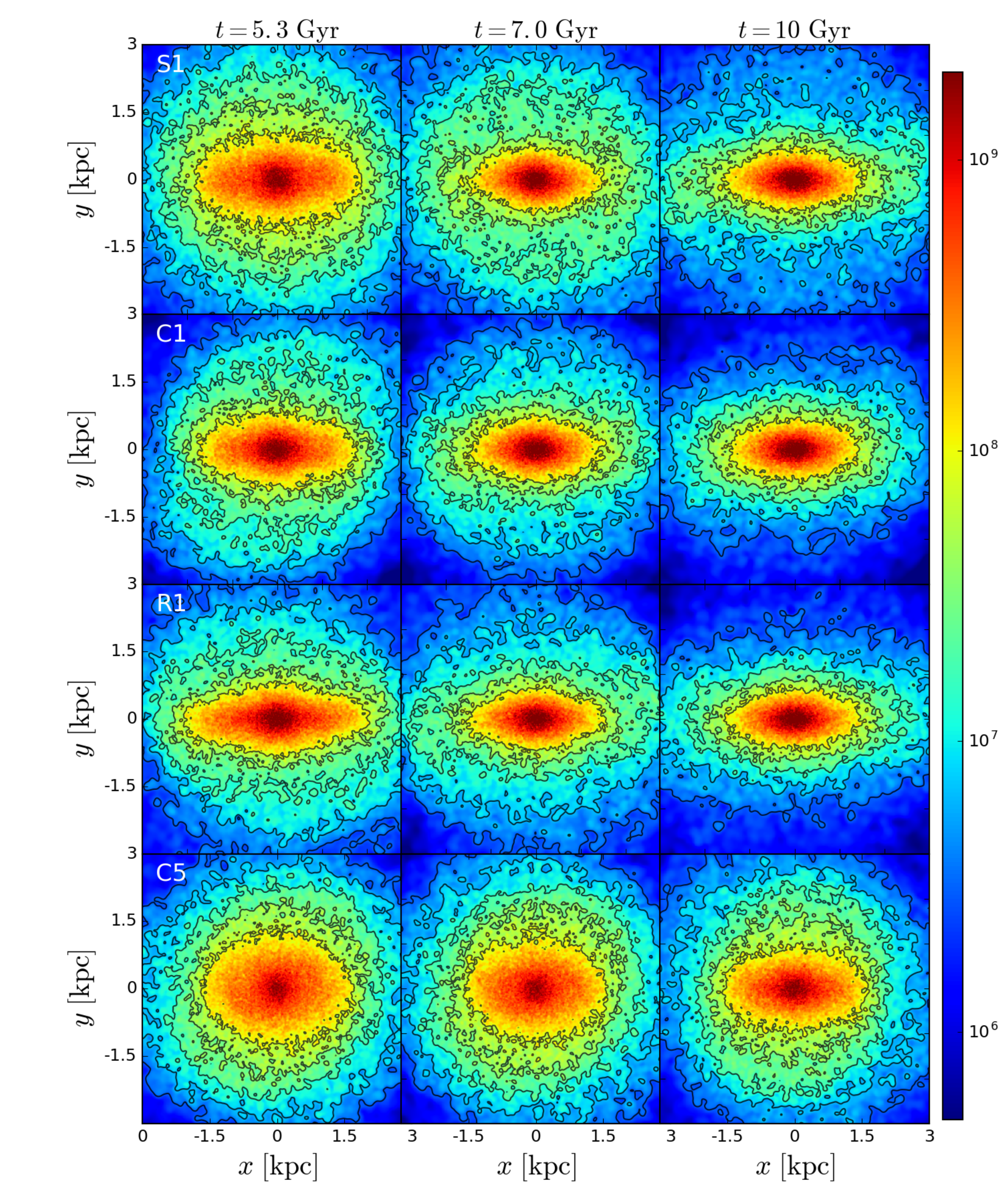}
\caption
{Face-on view of the stellar density $\rho$ in the $z = 0$ plane at three selected epochs, pre-buckling ($t=5.3\Gyr$), post-buckling ($t=7\Gyr$), and end of the run ($t=10\Gyr$) from left to right, for Galaxy S1 and Models C1, R1, and C5 from top to bottom. The colorbar gives $\rho$ in units of ${\rm\,M_\odot}{\rm\,kpc}^{-3}$.}\label{fig04face}
\end{figure*}

\begin{figure*}
\centering\includegraphics[angle=0,width=17cm]{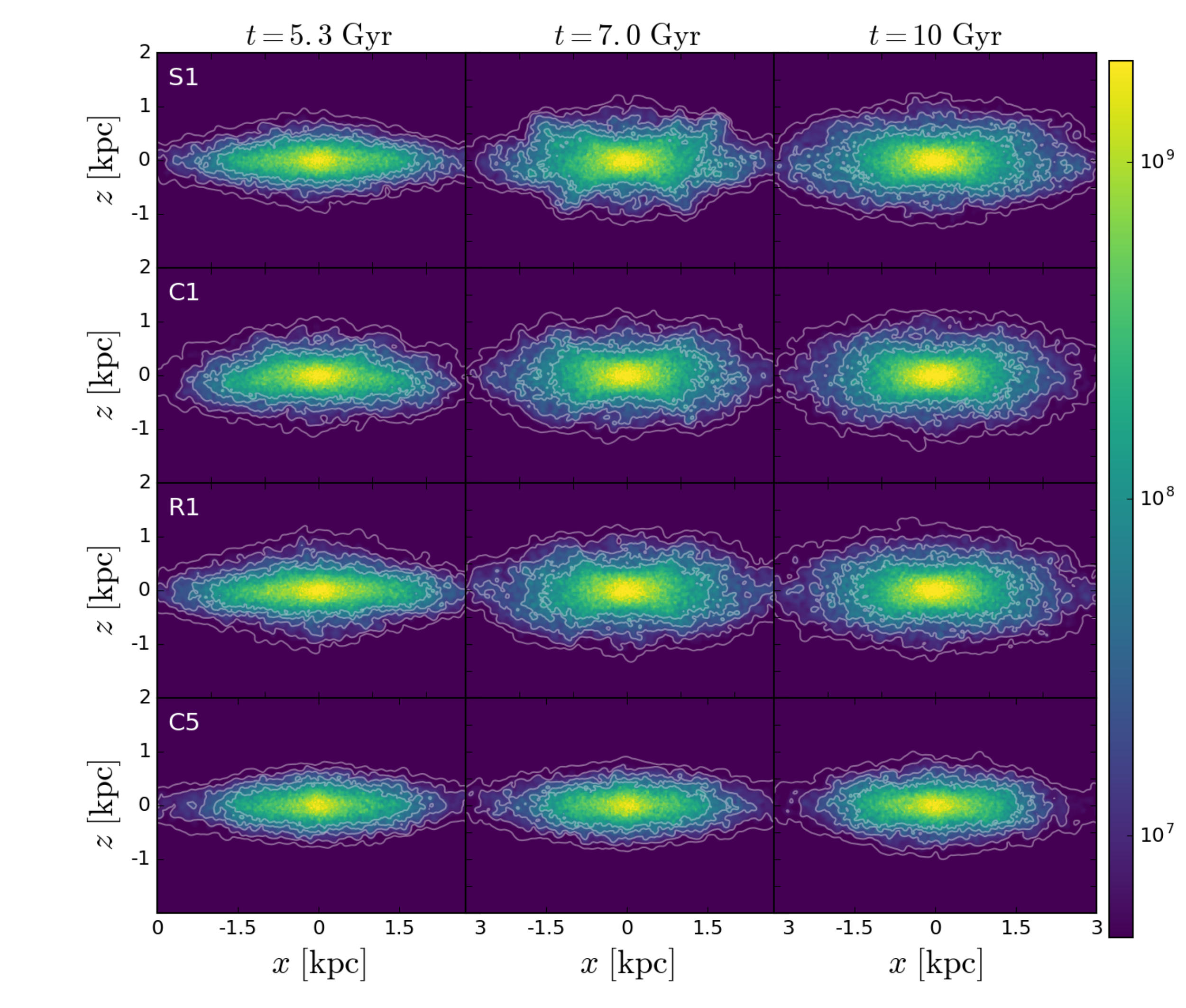}
\caption
{Edge-on view of the stellar density $\rho$ in the $y = 0$ plane at three selected epochs, pre-buckling ($t=5.3\Gyr$), post-buckling ($t=7\Gyr$), and end of the run ($t=10\Gyr$) from left to right, for Galaxy S1 and Models C1, R1, and C5 from top to bottom. The colorbar gives $\rho$ in units of ${\rm\,M_\odot}{\rm\,kpc}^{-3}$.}\label{fig05side}
\end{figure*}

\begin{figure}
\centering\includegraphics[angle=0,width=8.5cm]{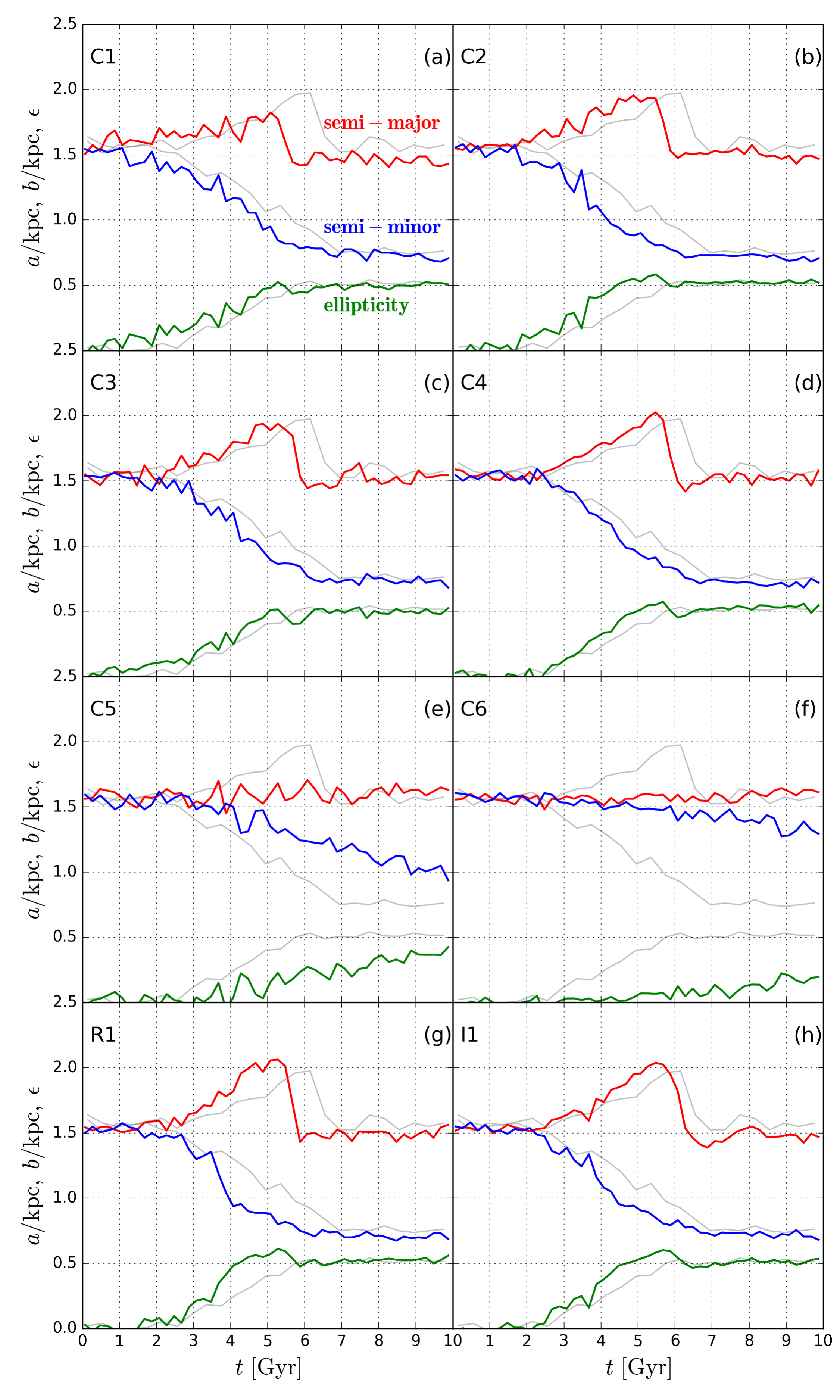}
\caption
{Temporal evolution of the semi-major axis $a$ (red), semi-minor axis $b$ (blue), and ellipticity $\epsilon=1-b/a$ (green) of the bars in all models. The results of Galaxy S1 in isolation are plotted in grey lines for comparison.}\label{fig06barlength}
\end{figure}

\begin{figure*}
\centering\includegraphics[angle=0,width=14cm]{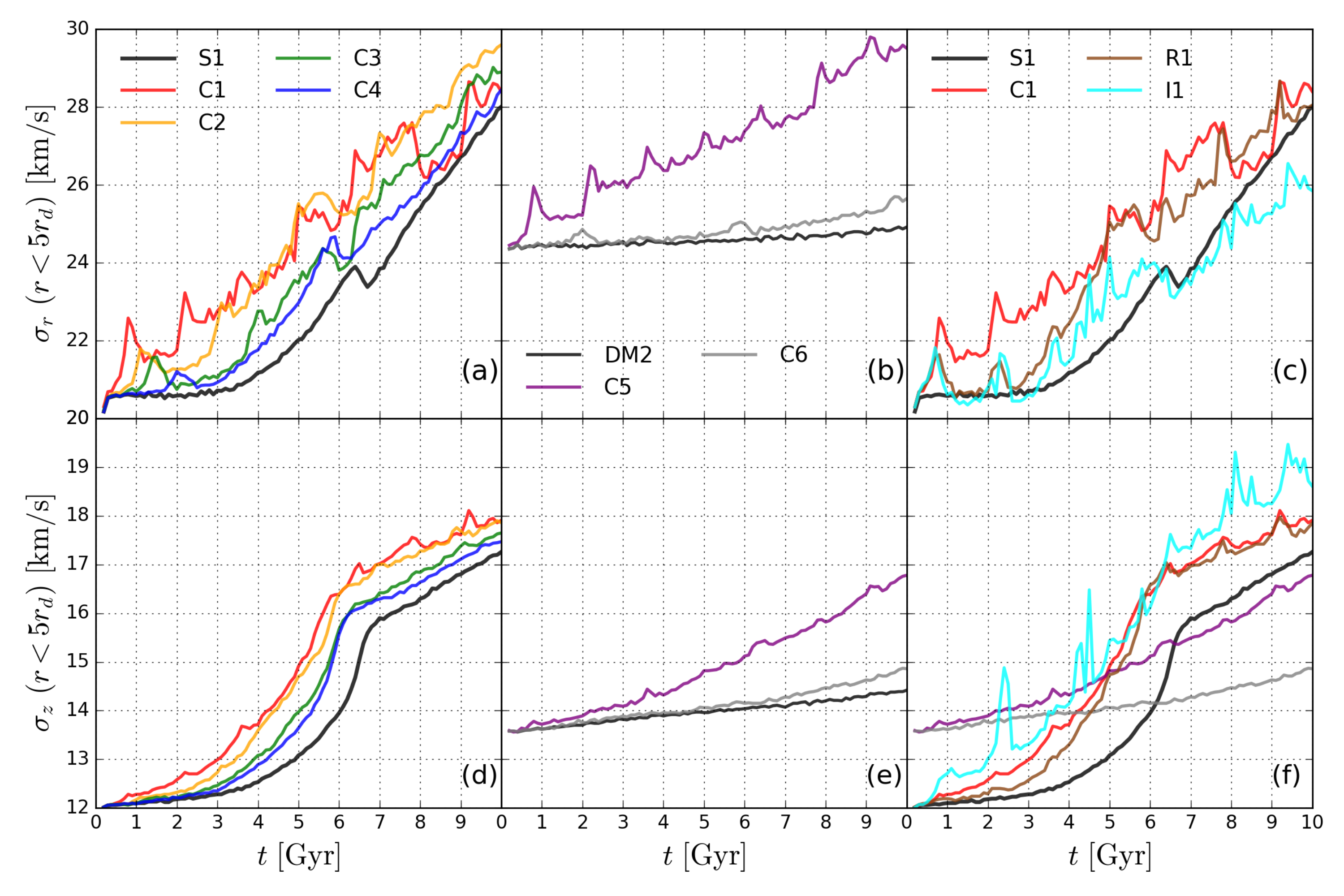}
\caption
{Evolution of the radial velocity dispersion $\sigma_r$ (upper panels) and the  vertical velocity dispersion $\sigma_z$ (lower panels) of the stellar component within $r=5r_d$ for all models. The results for Galaxies S1 and DM2 in isolation are compared as black solid lines.}\label{fig07disp}
\end{figure*}

\begin{figure}
\centering\includegraphics[angle=0,width=8.5cm]{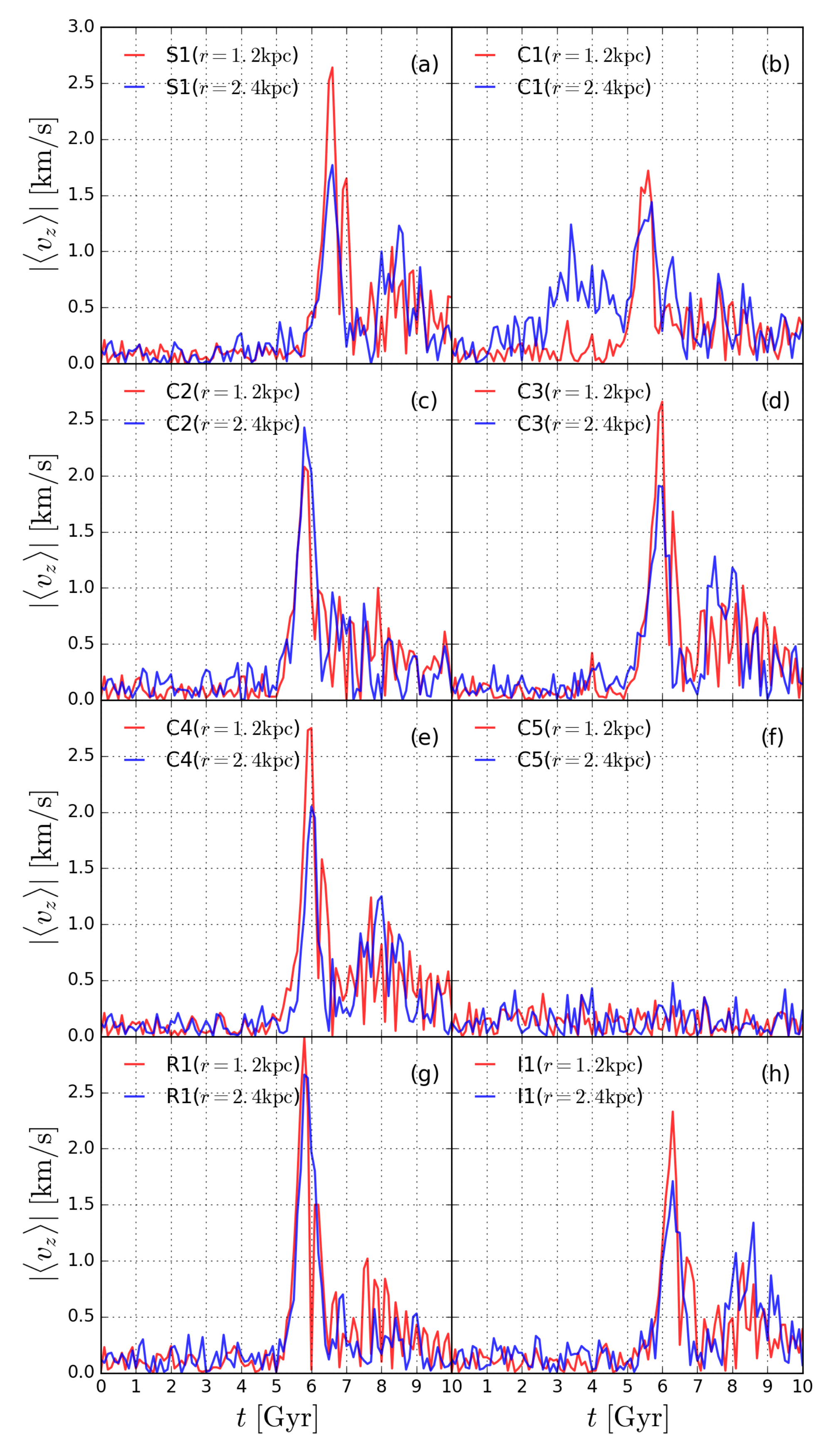}
\caption
{Temporal evolution of the mean vertical velocity $\vz$ at $r=1.2\kpc$ (red) and $r=2.4\kpc$ (blue) of the stellar component in Models C1--C5, R1, and I1. For comparison, the results for Galaxy S1 in isolation are plotted in (a).}\label{fig08vz}
\end{figure}

Tidal forcing not only produces spirals at early time (see Fig.~\ref{fig02spr}) but also excites stellar motions that are organized into a bar at late time.
Figure \ref{fig03fourier} plots the temporal evolution of the bar strength
$\langle F_2/F_0\rangle_\text{bar}$ averaged over $r=1.3$--$1.7\kpc$
in (a) Models C1--C4, (b) Models C5 and C6, and (c) Models C1, R1 and I1 with different spin orientations. For comparison, the results for Galaxies S1 and DM2 in isolation are drawn as the black line in each panel. The transition from the spiral-dominated disk to the bar-dominated disk occurs smoothly around $t=1$--$2\Gyr$.
The tidal forcing tends to make the bars form earlier by $\sim1$--$1.5\Gyr$ in Models C1--C4, compared to Galaxy S1 in isolation. However, stronger tidal force does not always correspond to faster and stronger bar formation, since a significant fraction of the tidal perturbations go to the spiral modes and heat the disk at early time.\footnote{The presence of spiral arms is known to increase the velocity dispersions of stellar particles (e.g., \citealt{sellwood84,carlberg85,jenkins90,sellwood02,desimone04,minchev06}).} Models C5 and C6 employing Galaxy DM2 with a compact dark halo do not possess a strong bar until the end of the run. Despite different spin orientations, the bar formation and growth in Models R1 and I1 are not much different from Model C1.

Figures \ref{fig04face} and \ref{fig05side} display the density slices of the stellar disks at $z=0$ and $y=0$ planes, respectively, for Models C1, R1, and C5, in comparison with Galaxy S1, at three epochs corresponding to the pre-buckling ($t=5.3\Gyr$), post-buckling ($t=7.0\Gyr$), and the final state ($t=10\Gyr$). The levels of the overlaid contours are in logarithmic scale between $10^{7.89}$ and $10^{6.5}\Msun \kpc^{-3}$ with the innermost level corresponding to the bar boundary. Galaxies in all models except for Models C5 and C6 undergo an episode of the buckling instability that weakens and thickens the bar, after which galaxy morphologies do not change much over time.
Even in Model I1, to which the tidal force is exerted vertically, the edge-on morphology is almost the same as the others.
This indicates that the bar formation and subsequent buckling instability are the major factors that alter the vertical structure, while the tidal effects by clusters are not significant on the morphological transformation.
Model C5 possesses a weak bar or an oval structure  until the end of the simulation, in which the velocity anisotropy is not high enough to trigger buckling instability.

We use the critical density $\rho_c=10^{7.89}\Msun$ kpc$^{-3}$ as the boundaries of the bars formed in our simulations.
Figure \ref{fig06barlength} plots the temporal changes of the bar semi-major axis $a$ (red), semi-minor axis $b$ (blue), and ellipticity $\epsilon=1-b/a$ (green). For comparison, the results of Galaxy S1 in isolation are plotted as the grey lines in all panels.
In Models C1--C4, the maximum value of $a$ is inversely proportional to $\TFC$. This is most likely due to tidal heating that causes the stellar orbits to deviate from bar-supporting orbits. Model R1 with the same $\TFC$ as Model C1 but has an inverted spin orientation relative to the orbit forms a longer bar than Model C1, suggesting that the bar growth is affected by the presence of spiral arms. On the other hand, Models C5 and C6 that host Galaxy DM2 do not form a clear bar but an oval with $\epsilon\lesssim 0.4$, suggesting tidal forcing with $\TFC\lesssim0.26$ is not strong enough to trigger bar formation in stable dwarf galaxies like Galaxy DM2.\footnote{\citet{gajda17} showed that a dwarf galaxy that is highly stable in isolation forms a bar subject to a sufficiently strong tidal force from a Milky Way-like host.}

The bar formation causes the orbits of stars to be elongated, increasing the radial velocity dispersion $\sigma_r$ until the end of the run (e.g., \citetalias{kwak17}).
In Figure \ref{fig07disp}, we plot the temporal evolution of the velocity dispersions within $r=5r_d$ of the stellar component in the (top) radial and (bottom) vertical directions.
Some spikes seen in the $\sigma_r$ curves for Models C1--C6 and R1 are induced by the impulsive tidal forcing near each pericenter passage that heats the disk temporarily. The corresponding changes in the $\sigma_z$ curves are relatively smooth, except for Model I1 in which the tidal forcing is significant along the $z$-direction near the pericenter approaches.
We note that a secular increase of $\sigma_r$ at $t\lesssim3\Gyr$ in Models C1--C4 in comparison with Galaxy S1 results from tidal forcing that forms spiral arms at early time, as shown in Figure \ref{fig01spr}.
Models R1 and I1 without spiral structure do not have such a secular increment of $\sigma_r$ at $t\lesssim3\Gyr$, but exhibit temporary peaks in $\sigma_r$, induced by the impulsive tidal forcing at pericenter passages (e.g., $t=0.70$ and $2.15\Gyr$).

A buckling instability occurs when the ratio of vertical to radial velocity dispersion decreases below a threshold value $\sigma_z/\sigma_r\lesssim0.3$ in a uniform disk \citep{toomre66,araki85}. For normal, radially-stratified disk galaxies, the critical values change to $\sigma_z/\sigma_r= 0.25\sim0.55$ in mid-disk regions and $0.66\sim0.77$ in the central regions \citep{raha91, sotnikova05, martinez06}.  \citetalias{kwak17} found that dwarf galaxies undergo a buckling instability if $\sigma_{z}/\sigma_{r}\lesssim 0.63$ at $r=0.9\kpc$. The bar formation and the ensuing angular momentum transfer, which increase velocity anisotropy by inducing radial random motions, are a favorable channel for the vertical buckling instability. Nevertheless, the presence of a bar is not a necessary condition for a vertical instability as \citetalias{kwak17} observed weak warping in the outer part of a non-barred dwarf galaxy, Galaxy DP4, because of its initially high radial velocity dispersion (see their Fig. 14(h)).

The occurrence and strength of the buckling instability are well recognized by means of the mean vertical velocity of stellar disks.
Figure \ref{fig08vz} plots the temporal variations of $\vz$ at $r=1.2\kpc$ (red) and $r=2.4\kpc$ (blue) for all models with a bar.
All models with a strong bar undergo the first episode of buckling instabilities, forming a sharp peak around $\sim5$--6$\Gyr$.
Models with lower $\TFC$ and longer bars suffer a stronger buckling instability, forming a larger peak in $\vz$ at $r=1.2\kpc$.
Similarly to Galaxy S1 in isolation, the secondary buckling instabilities follow afterward as a result of bar regrowth (\citealt{martinez06}; \citetalias{kwak17}).
The heights and epochs of the peaks in $\vz$ are slightly different owing to the different bar properties induced by the tidal force.
Model C1 also experiences vertical velocity asymmetry at $r=2.4\kpc$ around $t= 2$--5$\Gyr$, which is not a consequence of bar-bending buckling instability since $\vz$ at $r=1.2\kpc$ remains close to zero. Rather, we attribute this to a weak buckling instability of the outer regions where radial heating by the spiral arms and bar increases
the radial velocity dispersion (Fig. \ref{fig07disp}(c)), eventually making  $\sigma_z/\sigma_r$ drop below the threshold value for the instability.

\begin{figure}
\centering\includegraphics[angle=0,width=8.5cm]{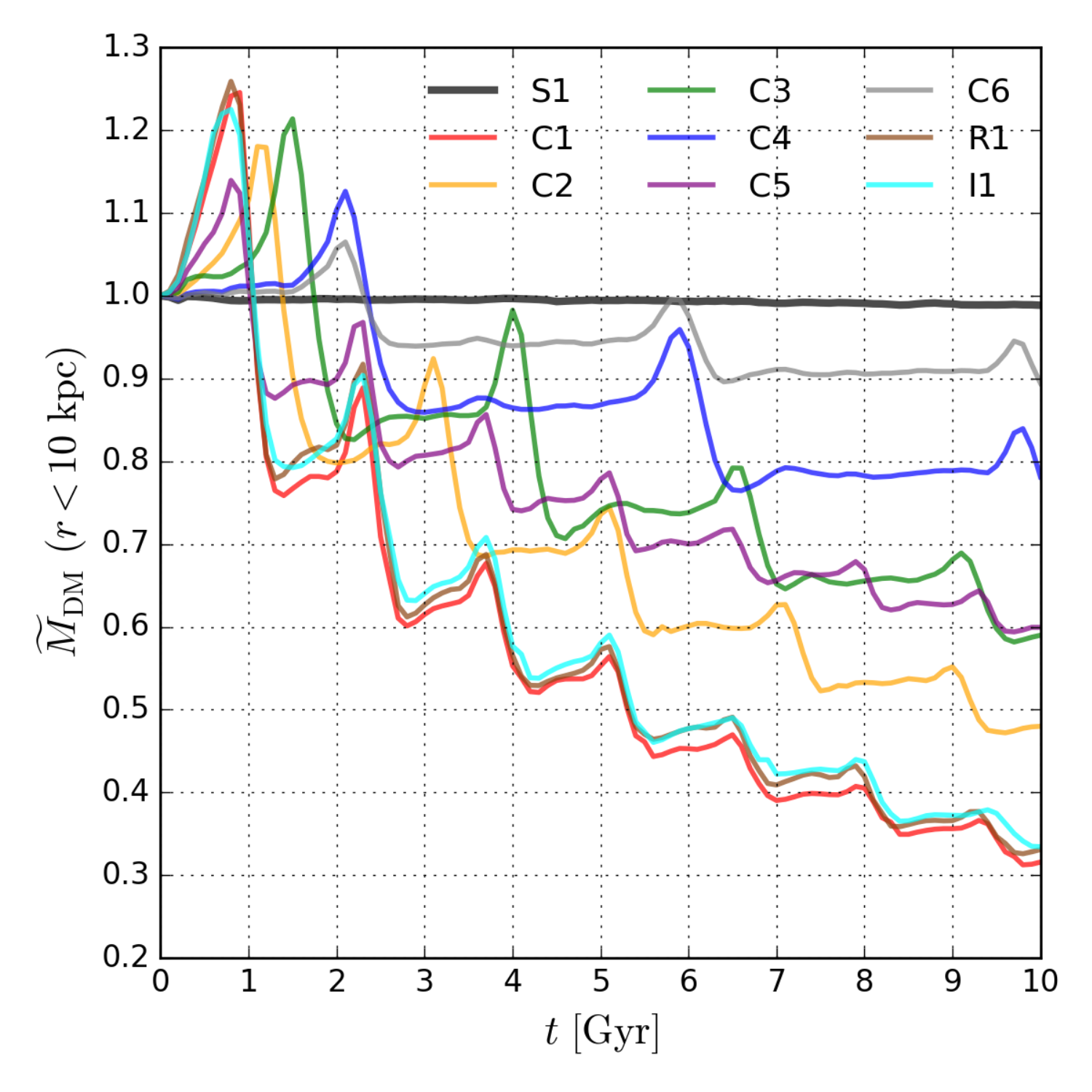}
\caption
{Evolution of the normalized enclosed mass of the galactic dark halo  $\normdm$ within $r=10\kpc$ relative to the initial value for all Cluster-Galaxy models. As the halo orbits about the cluster center, it undergoes tidal compression and stretching repeatedly before and after the pericenter passage, respectively, increasing and decreasing $\normdm$.}
\label{fig09dm}
\end{figure}

\begin{figure}
\centering\includegraphics[angle=0,width=8.5cm]{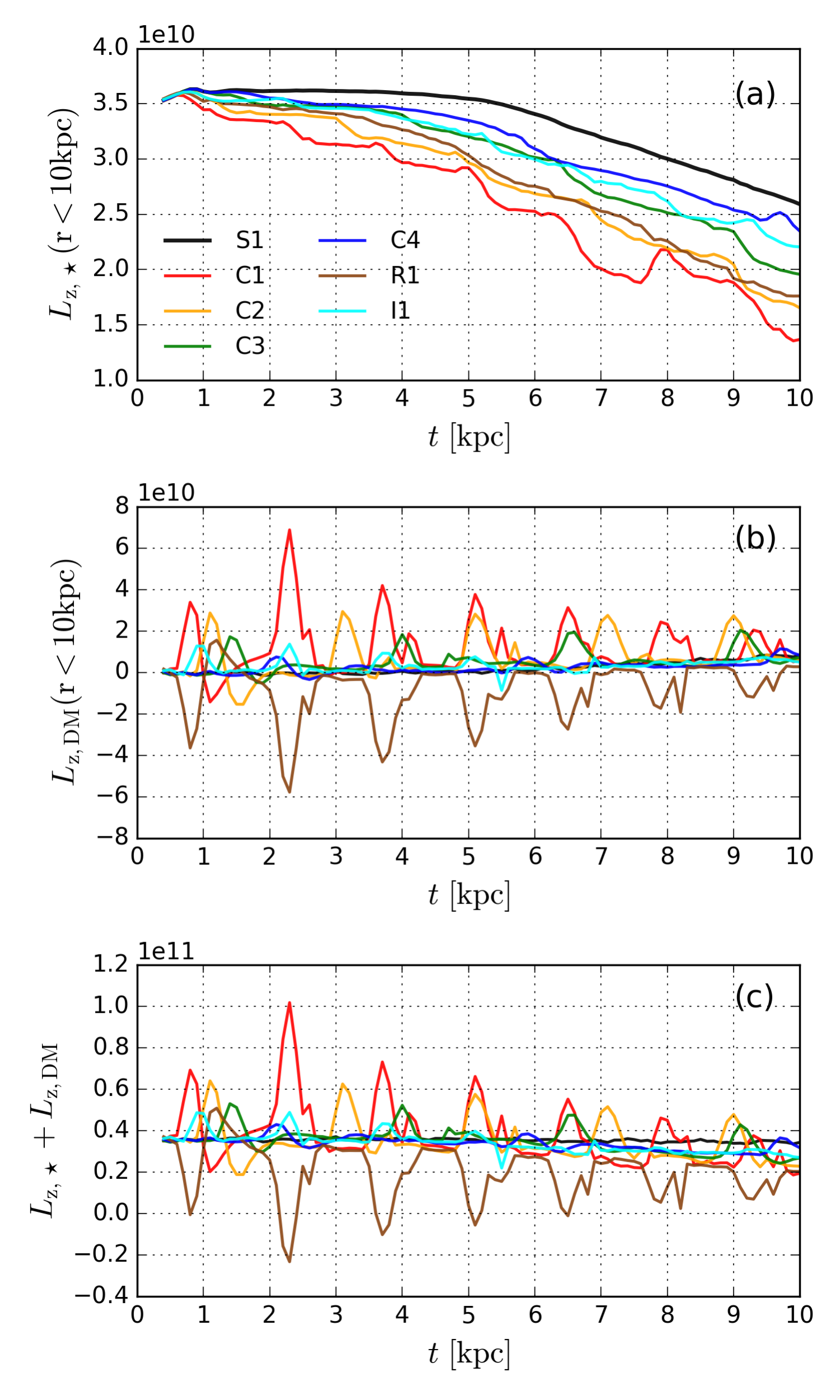}
\caption
{Evolution of the angular momentum in (a) the stellar disk within $r=10\kpc$, (b) the galactic halo within $r=10\kpc$, and (c) both disk and halo within the virial radius $r_\text{vir}=40\kpc$, in units of $\Msun\,\kpc\, \kms$.}\label{fig10angmom}
\end{figure}

\begin{figure}
\centering\includegraphics[angle=0,width=8.5cm]{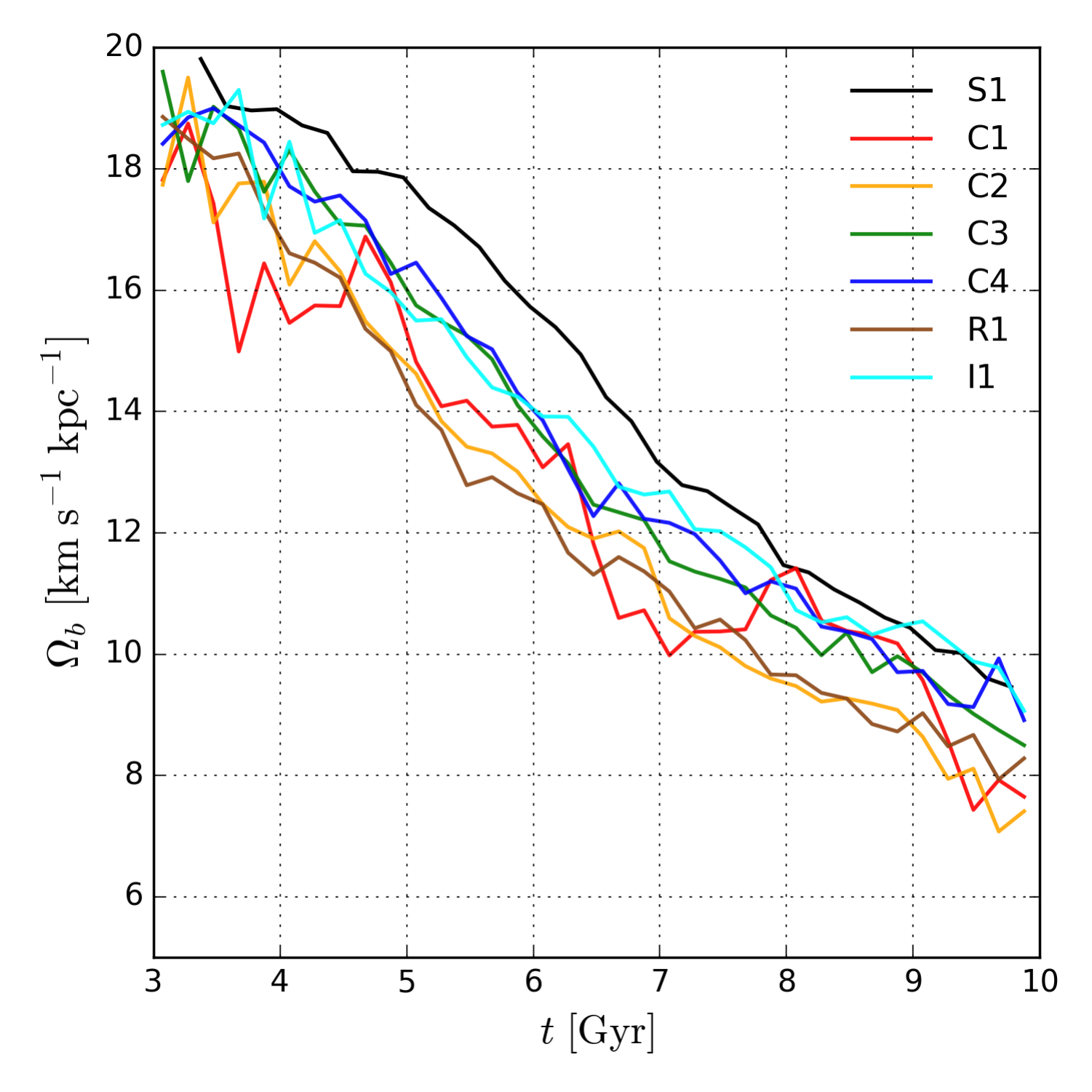}
\caption
{Evolution of the bar pattern speed $\Omega_b$ in all models with a bar. All bars slow down with time by transferring their angular momenta to the surrounding halos, with the rates insensitive to the strength of the tidal forcing.}\label{fig11omega}
\end{figure}

\subsubsection{Change in Mass and Angular Momentum}\label{s:ch3.2.3}
Owing to the tidal field in a cluster, galactic halos undergo a significant mass loss especially in the outer regions \citep{richstone76,white76,merritt83}.
Figure \ref{fig09dm} plots the temporal evolution of the enclosed mass of the  dark halo $\normdm$ within $r=10\kpc$ relative to the initial value for all models.
At early time, $\normdm$ increases temporally due to tidal compression as the galaxies approach the cluster center, and then starts to decrease due to tidal stretching as well as truncation right after passing the pericenter. This tidal compression and stretching pattern repeats at each pericenter passage.
The decrease in $\normdm$ is more vigorous in models with smaller $R_\text{peri}$ and a lower concentration parameter.
At $t=10\Gyr$, for instance, $\normdm(r<10\kpc)$ decreases by $\sim70\%$ and 10\% in Models C1 and C6, respectively.
The change of $\normdm$ within the initial virial radius is $\sim80\%$ at the end of the run for Model C1.
We also calculate the normalized enclosed mass of the stellar component, $\normstar$, within $r=10\kpc$ and find a similar temporal trend, although the decrease in $\normstar(r<10\kpc)$ is only $\sim4\%$ for Model C1.
The changes in $\normstar (t)$ and $\normdm (t)$ due to different spin orientations are negligibly small.

\citet{athanassoula03} demonstrated that a stellar bar continuously transfers the angular momentum to its live halo, and this angular momentum transfer is integral for the bar growth, slowdown, and buckling instability (see also \citealt{athanassoula05,berentzen06,athanassoula07,athanassoula14,sellwood16}; \citetalias{kwak17}). Figure \ref{fig10angmom} plots the temporal evolution of (a) the angular momentum $L_{z,\star}$ in the stellar disk inside $r=10\kpc$ and (b) angular momentum $L_{z,\text{DM}}$ in the galactic halo inside $r=10\kpc$, and (c) the combined angular momentum within the virial radius $r_{\rm{vir}}=40\kpc$ of the galactic halos for all models with Galaxy S1. Soon after the bar formation, all stellar disks begin to lose their angular momenta to the surrounding halos. While Galaxy S1 in isolation loses $\sim27\%$ of its initial $\lz_{,\star}$ to its halo in $10\Gyr$, the amount of the angular momentum loss increases to 34 and 62\% in Models C4 and C1, respectively, indicating that the tidal forcing promotes the angular momentum transfer. Among the models with different spin orientations, the angular momentum loss is lowest in Model I1, while Model R1 still loses $\sim50\%$ of the initial $\lz_{,\star}$ despite its inverted spin direction.

The evolution of $L_{z,\text{DM}}(r<10\kpc)$ is characterized by multiple peaks combined with a secular increase. The multiple peaks are caused by tidal torques at the pericenter passages, which occur due to misalignment between the major axis of the tidally distorted halo and the line connecting the halo and the cluster center (e.g., \citealt{pereira10}). In Model C1 (R1), the major axis lags (leads) the line connecting the halo and cluster center, resulting in positive (negative) tidal torque. Other than these peaks, $L_{z,\text{DM}}(r<10\kpc)$ secularly increases for all models as the halos absorb positive angular momenta from the disks.
The total angular momentum of Galaxy S1 is conserved as the galactic halo absorbs the amount of angular momentum lost from its rotating stellar bar (see also Fig.~6 in \citetalias{kwak17}). Due to the mass loss at consecutive pericenter passages, however, our tidal models do not conserve the total angular momentum within the initial virial radius of the galactic halos.

Figure \ref{fig11omega} plots the temporal changes of the bar pattern speed $\Omega_b$ for all models with a bar, showing that all bars slow down over time at similar rates. \citet{athanassoula03} showed that a stronger and longer bar tends to slow down more rapidly for galaxies in isolation (see also \citetalias{kwak17}). For normal galaxies in tidal interactions, on the other hand, \citet{lokas16} showed that stronger tidal forcing results in a stronger bar with lower $\Omega_b$. However, the tidal force by the cluster on our current models for dEdis is not much effective on $\Omega_b$ and its decay. In fact, the insensitivity of $\Omega_b$ to the tidal forcing is consistent with the results of \citet{miwa98} who found that the bar properties induced by a weak tidal force depend rather strongly on the intrinsic properties of a galaxy. A strong tidal force would instead `wash out' the intrinsic galaxy properties, making the bar properties dependent upon the tidal parameters. We note that $\Omega_b$ and $\lz_{,\star}$ in Model C1 simultaneously decrease at $t\sim6.35$ and $9.15\Gyr$ and increase at $t\sim7.75\Gyr$. This occurs because the bar can either speed up or slow down due to tidal torque at each pericenter passage, depending on its orientation relative to the line connecting the galaxy and cluster centers, as reported in \citet{lokas14,lokas16}.

Bars are considered to be slow if $\mathcal{R}\equiv r_\text{CR}/a$ is larger than $1.4$ \citep{elmegreen1996}, where $r_\text{CR}$ denotes the corotation resonance. For Models C1 and C4, $r_\text{CR}=2.98$ and 2.65 corresponding to $\mathcal{R}=1.32$ and  $1.51$, respectively, at $t=5.3\Gyr$ before the onset of the buckling instability.
These values increase to $\mathcal{R}=2.62$ and $2.66$ for Models C1 and C4, respectively, at $t=10\Gyr$ owing to the bar slowdown, suggesting that the bars formed in our models are fast at their formation, and subsequently lose angular momentum to turn to slow rotators (e.g., \citealt{seo19}).

Figure \ref{fig06barlength} shows that Model R1 forms a longer bar than Model C1 despite the same $\TFC$. Owing to its retrograde spin, Model R1 does not form any spiral structure that induces radial heating, which eventually hinders the bar growth.
Meanwhile, Model R1 still loses a large amount of $L_{z,\star}$ comparable to Models C1 and C2 at $t=5\Gyr$.
When the tidal force is not strong enough to wash out the intrinsic galaxy properties \citep{miwa98}, the maximum bar length achieved can be affected by the spiral-driven radial heating at early time as well as the amount of angular momentum transfer.

\section{Galaxy-Galaxy Interaction}\label{s:ch4}

\begin{deluxetable}{cccll}
\tablecaption{Galaxy-Galaxy Model Parameters\label{tbl:galgal}}
\tablewidth{0pt}
\tablehead{\colhead{Model}
          & \colhead{Galaxies}
          & \colhead{Pericenter}
          & \colhead{$\rm{TF_{G}}$}\\
            \colhead{$~$}
          & \colhead{Target, Perturber}
          & \colhead{[kpc]}
          & \colhead{$~$} \\
            \colhead{(1)}
          & \colhead{(2)}
          & \colhead{(3)}
          & \colhead{(4)}}
\startdata
G1  &  S1, DM2 & 9.6  & 0.26 \\
G2  &  S1, DM2 & 11.9  & {0.15} \\
G3  &  S1, DM2 & 14.2 & {0.095} \\
G4  &  S1, DM2 & 16.5 & {0.064} \\
G5  & DM2, DM2 & 9.6  & {0.16} \\
G6  & DM2, DM2 & 11.9  & {0.092} \\
G7  & DM2, DM2 & 14.2 & {0.059} \\
G8  & DM2, DM2 & 16.5 & {0.039} \\\enddata
\tablecomments{Column (1) is the model name. Column (2) lists the names of target and perturbing galaxies. Column (3) is the pericenter distance. Column (4) lists the dimensionless tidal force at the pericenter.}
\end{deluxetable}

\begin{figure}
\centering\includegraphics[angle=0,width=8.5cm]{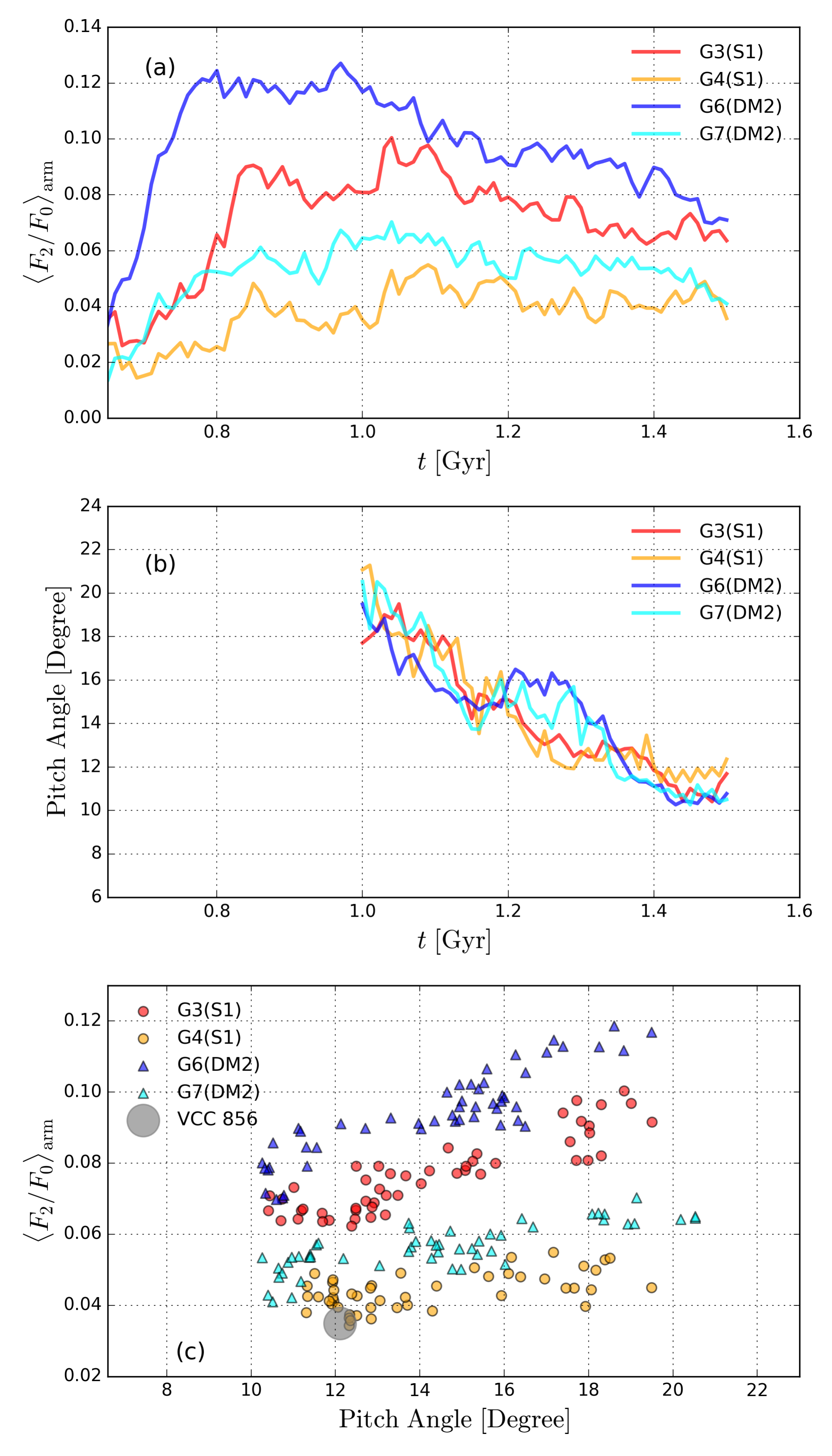}
\caption
{Temporal changes of (a) the strength $\langle F_{2}/F_{0} \rangle_\text{arm}$ and (b) the arm pitch angle $i$ of the arms measured at $r=1.0$--$2.0\kpc$ in Models G3, G4, G6, and G7. (c) Correlation between $\langle F_{2}/F_{0} \rangle$ and $i$ for Models G3, G4, G6, and G7. The properties of the arms in VCC 856 are marked by a grey circle.}\label{fig12summary}
\end{figure}

\begin{figure*}
\centering\includegraphics[angle=0,width=17cm]{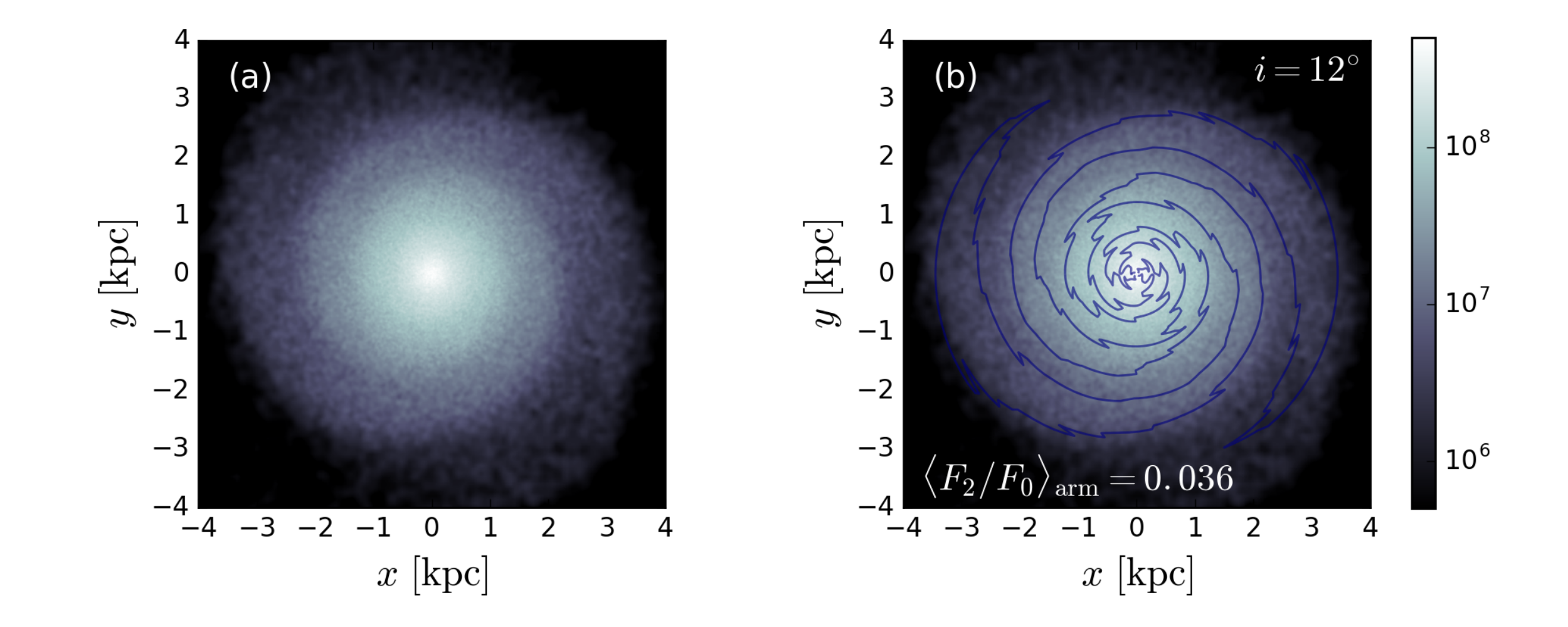}
\caption
{Face-on views of (a) the stellar density $\rho$ in the $z=0$ plane and (b) the overdensity contours of $F_2$ overlaid in Model G4 at $t=1.5\Gyr$ that resemble the arms in VCC 856. The colorbar gives $\rho$ in units of $\Msun\,\kpc\, \kms$. The arms have the strength $\langle F_2/F_0\rangle_\text{arm}=0.036$ and pitch angle $i=12^\circ$ at this time as indicated.}\label{fig13spr3}
\end{figure*}

The presence of spiral substructures with pitch angles similar to those of late-type galaxies has been frequently reported in many dEs, including VCC856 in the Virgo cluster (e.g., \citealt{jerjen00, lisker06a, lisker09, lisker09b}). As presented in Section \ref{sec:sparm}, our Cluster-Galaxy models form $m$=2 spirals driven by the cluster potential in the early phase of evolution. Such spirals have the arm strengths $\langle F_2/F_0 \rangle \sim 0.05$--$0.09$ and the pitch angles $i \sim 15.1^\circ$--$16.6^\circ$, which appear slightly more loosely wound than the observed arms in VCC856 with $i=12.1^\circ$.

With an aim to find probable conditions for the formation of faint and tightly wound spirals, we consider 8 models of Galaxy-Galaxy interactions on mutual prograding orbits, listed in Table \ref{tbl:galgal}. Models G1--G4 study interactions between Galaxies S1 and DM2, while Models G5--G8 consider interactions between two identical Galaxies DM2. As a measure of the tidal force in the galaxy-galaxy interactions, we define the dimensionless parameter $\TFS$ defined as
\begin{equation}
{\TFS}=\left(\frac{M_\text{ptb}}{M_\text{gal}}\right)
       \left(\frac{r_\text{gal}}{r_\text{peri}}\right)^{3},
\end{equation}
where $M_\text{ptb}$ is the total mass of the perturbing galaxy within $r_\text{peri}$, while $M_\text{gal}$ is the total mass of the target galaxy within $r_\text{gal}=7r_d$. The pericenter passage occurs at $t\sim0.65\Gyr$, and the corresponding $\TFS$ values for all models are listed in Column 4 of Table \ref{tbl:galgal}.

We present the simulation results from the Galaxy-Galaxy models only up to $t=1.5\Gyr$: the ensuing evolution is dominated by bar formation and an eventual galaxy merger even before a bar fully grows. The quasi-resonant radii are 3.1, 3.8, 4.4, and $5.0\kpc$ for models with the pericenter distance of $r_\text{peri}=9.6, 11.9, 14.2$, and $16.4\kpc$, respectively, which is passed at $t=0.65\Gyr$. Since the enclosed disk mass within $r=3.1\kpc$ is 96\% of the total, the galaxy-galaxy interactions are expected to be only moderate. Also, as discussed in Section \ref{s:ch3}, the spiral arms in Models G5--G8 that host Galaxy DM2 are expected to be fainter as their highly concentrated halos stabilize the disk.

At $t=1.5\Gyr$, Models G1, G2, and G5 with relatively strong tidal forcing are already dominated by a large-scale bar in the regions with $r\lesssim 1.5\kpc$, so that  their spiral-only phase lasts very briefly. In case of Model G8, The tidal forcing is too weak to produce spiral arms in the inner regions. It is only in Models G3, G4, G6 and G7 where we observe faint spiral arms without a long bar until $t=1.5\Gyr$. Figure \ref{fig12summary}(a) and (b) plot the temporal variations of the arm strength $\langle F_2/F_0\rangle_{\rm{arm}}$ and the arm pitch angle $i$ in the regions with $1.0\kpc \leq r\leq 2\kpc$. Similarly to the Cluster-Galaxy models, the arms in the Galaxy-Galaxy models decay and wind over time, but at a slightly faster rate, suggesting that they are kinematic density waves. This is presumably because the tidal force decays rapidly as the companion moves away from the pericenter in the latter models, while the change of the tidal force along the galaxy orbit is moderate in the former models. Again, the decaying rate of the arms is proportional to the arm strength. The arms in Model G7 are weaker than in Model G3 with the same $R_{\rm{peri}}$ due to the stabilizing effect of the concentrated halo.

Figure \ref{fig12summary}(c) plots $\langle F_2/F_0\rangle_{\rm{arm}}$ as a function of $i$ for Models G3, G4, G6, and G7.
The arms decay as they wind out, showing a weak positive correlation between the arm strength and the pitch angle.
The grey circle corresponds to the faint arms in VCC856, in excellent agreement with the arms in Model G4 at $t\sim1.4$--$1.5\Gyr$.
This suggests that VCC856 can be well described by Galaxy S1 that underwent tidal interactions of strength $\TFS=0.015$ with its neighbor about $\sim0.85\Gyr$ ago.  Figure \ref{fig13spr3} plots a face-on view of the stellar surface density and the spiral structures in the $z=0$ plane at $t=1.5\Gyr$ in Model G4.  Although the spiral arms are not clearly recognized from the surface density map especially in the inner regions, they manifest themselves as two-armed spirals in the distribution of $F_2/F_0$.

\section{Discussion and Summary}\label{s:ch5}

Using $N$-body simulations, we have examined the effects of the tidal forces on the evolution of dwarf galaxies. Our work begins from the assumption that VCC 856 is an infalling progenitor of disk-type dwarf galaxyies, which lacks gas contents presumably by ram-pressure stripping, but still preserves the dynamical properties from its early phase without undergoing mergers or strong tidal encounters. \citetalias{kwak17} investigated the stability of VCC 856 in isolation by varying parameters within the observed error ranges, finding that the infalling progenitors are intrinsically unstable to the bar formation.
As the second paper of the series, we study the tidal effects by either cluster potential or a companion galaxy on the formation of bars and spirals. We
consider two types of models: Cluster-Galaxy models and Galaxy-Galaxy models. In the Cluster-Galaxy models, we place two galaxy models (S1 and DM2) taken from \citetalias{kwak17} around an NFW halo whose mass, size, and concentration are similar to the Virgo cluster. Galaxies S1 and DM2 are nearly identical except that the latter has the dark matter fraction within the effective radius about twice higher than the former. In the Galaxy-Galaxy models, we let two galaxies undergo tidal encounters by differing the pericenter distance. It should be reminded that VCC 856 is a bulgeless galaxy and has spiral arms with pitch angle $i=12^\circ$ and the fractional arm amplitude $\sim3$--$4\%$ in the inner disk.

\subsection{Cluster Effects}
We find that the tidal effects by the Virgo-like cluster halo on the stability of dwarf disk galaxies are moderate.
In our Cluster-Galaxy models with Galaxy S1, the bar formation occurs earlier by only $\sim1\Gyr$ than in Galaxy S1 in isolation (Fig.~\ref{fig03fourier}).
The models with different spin orientations also form a bar at nearly the same epoch as the prograding counterpart. On the other hand, the tidal force that forms a bar in Models C1 to C4 does not trigger the bar formation in Models C5 and C6 with Galaxy DM2.  This indicates that the effect of the cluster tidal force on the bar formation in dwarf disk galaxies is not significant. The main reason for this is of course dwarf galaxies are small in size and mass.

Owing to the weak tidal forces, the properties of the bars such as length, shape (Fig.~\ref{fig06barlength}), and pattern speed (Fig.~\ref{fig11omega}) are not much different from those of the bars formed in isolation.
This is consistent with \citet{miwa98} who suggested that the properties of bars by a weak tidal force highly depend on the kinematics of galaxies, while a strong tidal encounter can significantly alter the outcomes.
Compared to \citet{lokas16} who studied evolution of normal disk galaxies in clusters with the same cluster halo and orbital parameters with our Models C1 to C4, the properties of the bar in their Model S1 on the innermost orbit are found to be strongly dependent on the tidal forces because their galaxy models are more massive with higher $\TFC$ than ours.
Especially, the initial pattern speed of their Model S1 is much lower, whereas the rest of their models that undergo relatively weak tidal forces have pattern speeds similar to those obtained in our paper.

Although the tidal force does not affect the bar formation much, it is able to induce spiral arms at early time (Fig.~\ref{fig02spr}) and subsequent radial heating (Fig.~\ref{fig07disp}) in the disks whose spin axis is parallel to the orbit axis. As in \citet{minchev06} who demonstrated that the formation of spirals is another heating mechanism aside from the bar formation, the disk heating in the radial direction is enhanced only in our prograding models with spiral structures, with the amount of heating proportional to the tidal force.
Meanwhile, the velocity anisotropy is directly related to the occurrence of buckling instability \citep{combes81,combes90,raha91,merritt94} and even vertical heating before the bar formation (\citetalias{kwak17}).
Model C1 that suffers the most radial heating on the innermost orbit is ended up undergoing continual disk warping in the vertical directions (Fig.~\ref{fig08vz}) as a result of increasing velocity anisotropy even before its bar fully grows. Amongst all models, Model C1 forms the shortest bar owing to the spiral-driven radial heating, the tidal heating, and the vertical warping, all of which eventually cause the stellar orbits to move away from bar-supporting orbits.

Whereas the stellar components of our models display no dramatic change in mass, the galactic halos experience vigorous mass stripping multiple times near the pericenter passages, losing up to $\sim80\%$ of the total mass within the initial virial radius (Fig.~\ref{fig09dm}). From the comparison between Models C1 and C5, the mass truncation is more effective on inner orbits and less effective in more compact galaxies, consistent with the previous results that the tidal truncation is moderate in more compact systems \citep{richstone76,white76,merritt83,gnedin03b}. This halo truncation can potentially reduce the stabilizing effect of the halo against the bar formation (\citealt{ostriker73,christodoulou95,sellwood01}; \citetalias{kwak17}).

The bars formed in our models slow down as they lose their angular momenta to the surrounding halos (Fig.~\ref{fig10angmom}), which in turn boosts the bar growth (\citealt{athanassoula05,berentzen06,athanassoula07,athanassoula14,sellwood16}; \citetalias{kwak17}). In case of the prograde spin (e.g., model C1), the cluster tidal force makes the bar rotate temporarily faster and even controls the bar orientation on the innermost orbit, taking away $\sim62\%$ of the initial $L_{z,\star}$ (e.g., \citealt{lokas16}). Even in the retrograding model, Model R1, we find that its bar still loses a comparable amount of $L_{z,\star}$.
However, unlike the prograding galaxies, the retrograding galaxy does not suffer from the spiral-driven radial heating, so that it ends up with possessing a longer bar than Model C1. All the bars formed in our models undergo vertical buckling instabilities, resulting in bar shortening and disk thickening (Fig.~\ref{fig05side}).
The strength of the vertical instability is proportional to the bar length and strength (Figs.~\ref{fig06barlength} and \ref{fig08vz}), consistent with the results of \citetalias{kwak17}.

We find that the infalling dwarf galaxies overall do not undergo any dramatic morphological transformation by the cluster tidal forces.
One major morphological transition is the episode of the buckling instability, but its occurrence is irrelevant to the cluster as it also arises in isolated dwarf disks galaxies. Hence, the formation of kinematically static dEs and lens-like components, which are known as `defunct bars' \citep{kormendy79,kormendy04}, is not observed in our simulations.
Considering that a realistic cluster is not isotropic and grows by mergers in time, the actual orbits of galaxies would be irregular \citep{gnedin03a}, and the cluster effects would be even weaker as the cluster mass would be lower in early time \citep{moore99}.
Yet, among other cluster environments, direct galaxy-galaxy collisions and multiple tidal shocking that are not considered in this work may happen about 10 times during the hierarchical cluster formation process \citep{gnedin03a} and even transform a large disk galaxy into an S0 galaxy \citep{gnedin03b}. We conjecture that such strong tidal forcing from high-speed encounters would be more severe for dwarf galaxies with lower kinetic energy.
We defer to the future work quantitative studies on the distribution of the barred populations and the formation of missing populations formed possibly by multiple, high-speed encounters.

\subsection{Formation of Faint Spirals}
In all the Cluster-Galaxy models, the tidal field of the cluster potential is moderate on dwarf galaxies and can trigger the formation of faint spirals that were unseen in simulated galaxies in isolation (e.g., \citetalias{kwak17}). A two-armed spiral structure forms only in models with prograding galaxies.
The arms are weaker in models with lower tidal strength and in more dark-matter dominant galaxies (Figs.~\ref{fig01spr} and \ref{fig02spr}). The sprial arms produced by the cluster tidal field are more loosely wound than those observed in VCC856.

In Galaxy-Galaxy models, two of our target galaxies (S1 and DM2) with the pericenter distance $r_{\rm{peri}}=9.6\kpc$ immediately result in the bar formation due to strong tidal forces.
On the other hand, Galaxy DM2 with $r_{\rm{peri}}=16.5\kpc$ does not form clear spirals in the disk, due to the weak tidal forces and the stabilizing effect of the compact halo.
Compared to the evolution of the spiral arms formed under the cluster tidal field, the spiral arms induced by its neighbor decay faster as the tidal forcing by its companion fade away rapidly.

Less than 1 Gyr after the galaxy interaction, Galaxy S1 in Model G4 forms tightly-wound and faint spiral arms similar to the observed ones in VCC 856.
\citet{jerjen00} suggested that the origin could be either swing amplification of internally-driven perturbations or a tidal force by a nearby galaxy, and our results advocate the tidal origin.
For the formation conditions of spiral arms, we emphasize that the tidal force should be weak, and the stellar disk needs to be marginally unstable without an excessively concentrated halo.
If the gravitational potential of the Virgo cluster is the driver for such spirals, VCC 856 should be on a prograding orbit with a tidal force somewhere between Models C3 and C4.

\acknowledgments
We thank the anonymous reviewer for his/her constructive suggestions. S.K. thanks In Sung Jang for useful discussions. S.K. and S.C.R appreciate Helmut Jerjen's helpful suggestions at the KASI-CNU Joint Workshop. The work of W.-T.~Kim was supported by the National Research Foundation of Korea (NRF) grant funded by the Korea government (MSIT) (2019R1A2C1004857). S.C.R. was supported by the Basic Science Research Program through the NRF of Korea funded by the Ministry of Education (2018R1A2B2006445). Support for this work was also provided by the NRF to the Center for Galaxy Evolution Research (2017R1A5A1070354). The computation of this work was supported by the Supercomputing Center/Korea Institute of Science and Technology Information with supercomputing resources including technical support (KSC-2018-CHA-0047).

\end{document}